\documentclass[12pt,twoside]{article}
\usepackage{amsmath}\usepackage{bbm}
\usepackage{dsfont}%\usepackage{authordate1-4}

\newcommand{\be}{\begin{equation}}\newcommand{\ee}{\end{equation}}
\newcommand{\bea}{\begin{eqnarray}}\newcommand{\eea}{\end{eqnarray}}
\newcommand{\nn}{\nonumber}
\newcommand{\pa}{\partial}
\newcommand{\ben}{\begin{enumerate}}\newcommand{\een}{\end{enumerate}}

\newcommand{\al}{\alpha}
\newcommand{\ga}{\gamma}
\newcommand{\ep}{\varepsilon}
\newcommand{\om}{\omega}\newcommand{\Om}{\Omega}
\newcommand{\tphi}{\tilde{\phi}}%\newcommand{\phit}{\tilde{\phi}_{\om}}
\newcommand{\txi}{\tilde{\xi}}\newcommand{\dxi}{\dot{\xi}}
\newcommand{\hxi}{\hat{\xi}}
\newcommand{\hq}{\hat{q}}

\newcommand{\sump}{{\sum\limits_{l=0}^{\infty}}{\vphantom{\sum}}^{\prime}}

\newcommand{\Ref}[1]{(\ref{#1})}
\renewcommand{\ni}{\noindent}

\begin{document}
\title{
Vacuum and thermal energies for two oscillators interacting through a field}
\author{M. Bordag\footnote{bordag@uni-leipzig.de}\\
\small Institut f\"ur Theoretische Physik, Universit\"at Leipzig, Germany.}
\date{}\maketitle
%\date{\today, file: \jobname}

%
\begin{abstract}
    We consider a simple (1+1)-dimensional model for the Casimir-Polder interaction consisting of two oscillators coupled to a scalar field. We include dissipation in a first principles approach by  allowing  the oscillators to interact with heat baths. For this system, we derive an expression for the free energy in terms of real frequencies. From this representation, we derive the Matsubara representation for the case with   dissipation.   Further we consider the case of vanishing intrinsic  frequencies of the oscillators. We show that in this case  the contribution from the zeroth Matsubara frequency gets modified and no  problems with the laws of thermodynamics appear.
\end{abstract}
%
%%%%%%%%%%%%%%%%%%%%%%%%%%%%%%%%%%%%%%%%%%%%%%%%%%%%%%%%%%%%%%%%%%%%%%%%%%%%%%%%
\section{\label{T0}Introduction}
The Casimir-Polder force describes the basic electromagnetic interaction between molecules or between a molecule and a wall, and the Lifshitz formula describes the interaction between macroscopic bodies. Their theoretical foundation is commonly based on stochastic electrodynamics and incorporates material properties in terms of permittivity or polarizability. This way, it is possible to account for the real structure of the interacting objects and to make predictions for high precision measurements, which constitute a highly actual and important topic.

In line with great success, there are some subtleties when accounting for dissipation.  In the simplest case, inserting Drude permittivity into the Lifshitz formula, a thermodynamic  principle was shown to be violated \cite{beze02-65-052113} and,  in another case, accounting  for dc-conductivity in atom-wall interactions, a   disagreement with experimental results was found \cite{klim08-41-312002}. For the current state of this problem see \cite{klim16-94-045404}. Also, there is the problem that a fixed permittivity, if going to infinity, does reproduce the ideal conductor limit only if using Schwinger's prescription \cite{schw78-115-1}.

It must be mentioned that insertion of some model for permittivity or polarizability in case of Casimir-Polder forces, into the Lifshitz formula, is a phenomenological approach and one may ask for a more fundamental one. As such, it is common to introduce the coup\-ling of the oscillators, describing the molecules,  to a heat bath (reservoir). Such an approach allows for a formulation from first principles, starting from a Hamiltonian, for instance. Then the 'remaining' task is to work out that for a specific configuration. For Casimir-Polder and Casimir forces such a system consists of dipoles interacting with the electromagnetic field and interacting with their respective heat baths. Thereby, the dissipation results from the interaction with the heat baths and is accompanied by Langevin forces.

There is a large number of papers implementing these ideas, for instance
\cite{kupi92-46-2286},\cite{rosa10-81-033812},\cite{rosa11-84-053813},\cite{lomb11-84-052517},\cite{berm14-89-022127},\cite{brau17-190-237},
to mention some of those which are related to Casimir-Polder and Casimir forces. Also, the approach with heat baths allows to consider non-equilibrium situations, radiative  heat transfer and a variety of further phenomena more generally appearing in  open quantum systems \cite{breuer2002}. However, the existing papers consider a homogeneous medium, are restricted to  oscillators with non zero intrinsic frequency or to the large separation expansion and reproduce mainly the known results on Casimir-Polder forces or Lifshitz formula.

In a different, but in some sense equivalent way, the problem may be formulated  in terms of frequencies. As known, in the case without dissipation, the considered problem has real eigenfrequencies and thermodynamic quantities like free energy may be represented as sum/integral over these frequencies. The equivalent representation in terms of imaginary Matsubara frequencies is related by a Wick rotation. In the case with dissipation, the eigenfrequencies are not real, and the free energy cannot be represented in those terms, but only in terms of Matsubara frequencies. The attempt to naively make a Wick rotation backwards will fail \cite{bord11-71-1788}. As it was shown in \cite{intr12-86-062517}, a representation in terms of complex frequencies (resonances) is possible, but not especially enlightening.

In the present paper, we work out the fundamental approach of describing dissipation by   coupling   to heat baths for the simplest system mimicking the Casimir-Polder interaction. We consider two one dimensional  harmonic oscillators, each coupled to a heat bath and both coupled to a (1+1)-dimensional scalar field. We do not make the large separation expansion from the beginning and have thus to include nonperturbative effects which modify the Matsubara representation. The oscillators have intrinsic frequency $\Om$, which may be put to zero during or at the end of the calculations. We express the free energy of this system as sum/integral over real frequencies and investigate the properties of this representation, for instance, we show the relation to the Matsubara representation and consider the limits of low temperature, vanishing dissipation and vanishing intrinsic frequency $\Om$.

The basic ideas of heat baths, used in this paper, are not new. These are basically the same as, for instance,  in the famous Fano paper \cite{Fano61-124-1866}, in the Huttner-Barnett approach \cite{hutt92-46-4306} and in the open quantum systems approach \cite{lomb11-84-052517}. However, in the context of Casimir and Casimir-Polder interactions with dissipation, these ideas became popular not so long ago. For this reason, we give here the most simple derivation which can be followed easily and  which is sufficient to describe just all necessary features we are interested in.
Also, it must be mentioned that the heat bath approach can be reformulated in an equilibrium situation in terms of the fluctuation-dissipation theorem, dating back to the work \cite{call51-83-34}.

We start in the next section with the interaction of a single oscillator with a heat bath. Then, we consider the interaction of two oscillators with the scalar field. In Section \ref{T3}, we consider  the complete system in a finite box and derive the representation in terms of real frequencies. In section \ref{T4},   we consider some special cases and the transition to Matsubara frequencies. In section \ref{T5}, we consider the system in unrestricted space and, in Section \ref{concl}, we draw conclusions.

\ni Throughout the paper we use units with $c=k_{\rm B}=1$.

\section{\label{T1}One oscillator interacting with a heat bath}
In this section, we consider a single harmonic oscillator interacting with a heat bath. This is one of the simplest thermodynamic systems and can be found in many textbooks. Here, we repeat the basic formulas in a way which is convenient for our further applications.
\subsection{\label{secT1.1}Basic formulas}
We consider a harmonic oscillator $\xi(t)$, coupled to an infinite collection of heat bath oscillators $q_\om(t)$. This system has a Lagrangian,
\be {\cal L}=
\frac{m}{2}\left(\dot{\xi}(t)^2-\Om^2\xi(t)^2\right)
   +\int_0^\infty d\om\,\frac{\mu}{2}\left(\dot{q}_\om(t)^2-\om^2\left(q_\om(t)-\xi(t)\right)^2\right),
\label{T1.1}\ee
where $\mu$ is the mass density of the bath oscillators to be specified below. The coupling is chosen after \cite{ford88-37-4419} in a way to avoid artificial divergences. In \Ref{T1.1}, all variables have time dependence, which we drop in the following to keep notations simple. The Hamiltonian corresponding to \Ref{T1.1} is
\be H=H_{\rm osc}+H_{\rm bath}+H_{\rm int}
\label{T1.2}\ee
with
\bea H_{\rm osc}&=&\frac{p^2}{2m}+\frac{m\Om^2}{2}\xi^2,\label{T1.3a}\\
    H_{\rm bath}&=&\int_0^\infty d\om\,\left(\frac{p_\om^2}{2\mu}+\frac{\mu\om^2}{2}q_\om^2\right),\label{T1.3b}\\
    H_{\rm int}&=&\int_0^\infty d\om\, \frac{\mu\om^2}{2}(-2q_\om+\xi)\xi,
\label{T1.3c}\eea
and the conjugated momenta are
\be p=\frac{\pa {\cal L}}{\pa \dot{\xi}},\quad
    p_\om=\frac{\pa {\cal L}}{\pa \dot{q}_\om}.
\label{T1.4}\ee
The equations of motion following from the Lagrangian \Ref{T1.1} are
\be m(\ddot{\xi}+\Om^2\xi)=\int_0^\infty d\om\, {\mu\om^2}(q_\om-\xi)
\label{T1.5}\ee
for the oscillator and
\be \mu(\ddot{q}_\om+\om^2q_\om)= {\mu\om^2}\xi
\label{T1.6}\ee
for the bath variable.

Next we quantize the system \Ref{T1.1} resp. \Ref{T1.2}. We introduce creation and annihilation operators $\hat{b}^\dagger_\om$ and $\hat{b}_\om$ for the bath with commutator relation
\be [\hat{b}_\om,\hat{b}^\dagger_{\om'}]=\delta(\om-\om')
\label{T1.7}\ee
and by means of
\be \hat{q}_\om=\frac{l_0}{\sqrt{2}}\left(\hat{b}_\om+\hat{b}^\dagger_\om\right),\ \
\hat{p}_\om=\frac{\hbar}{ i \sqrt{2}l_0}\left(\hat{b}_\om-\hat{b}^\dagger_\om\right),
\label{T1.8}\ee
the corresponding operators for  coordinate and momentum. Here
\be l_0=\sqrt{\frac{\hbar}{\mu \om}}
\label{T1.9}\ee
is the characteristic length associated with the bath oscillators.

The Hamilton operator associated with \Ref{T1.3b} is
\be \hat{H}_{\rm bath}=\int_0^\infty d\om\,\frac{\hbar\om}{2}(\hat{b}^\dagger_\om\hat{b}_\om+\hat{b}_\om\hat{b}^\dagger_\om).
\label{T1.10}\ee
Next we consider the Heisenberg equations of motion for the operators $\hat{b}^\dagger_\om(t)$ and $\hat{b}_\om(t)$ (now we show the time dependence again),
\bea    \dot{\hat{b}}_\om(t)&=&\frac{i}{\hbar}[\hat{H},\hat{b}_\om(t)]=-i\om\hat{b}_\om(t)+i\frac{\om}{\sqrt{2}l_0}\hat{\xi}(t),
\nn\\
 \dot{\hat{b}}^\dagger_\om(t)&=&\frac{i}{\hbar}[\hat{H},\hat{b}^\dagger_\om(t)]=i\om\hat{b}^\dagger_\om(t)-i\frac{\om}{\sqrt{2}l_0}\hat{\xi}(t),
\label{T1.11}\eea
where \Ref{T1.9} was used, and with \Ref{T1.8} we get the equation for the oscillator,
\be  m(\ddot{\hat{\xi}}+\Om^2\hat{\xi})
        =
        \int_0^\infty d\om\, \frac{\mu\om^2l_0}{\sqrt{2}}(\hat{b}_\om(t)+
        \hat{b}^\dagger_\om(t))
-\int_0^\infty d\om\,\mu\om^2\hat{\xi}(t).
\label{T1.12}\ee
The solutions to eqs. \Ref{T1.11} are the Heisenberg operators,
\bea \hat{b}_\om(t)&=& e^{-i\om t}\hat{b}_\om(0)
-\frac{i\om}{\sqrt{2}l_0}\int_{-\infty}^t dt'\,
e^{-i\om(t-t')}\hat{\xi}(t'),\nn\\
 \hat{b}^\dagger_\om(t)&=& e^{i\om t}\hat{b}^\dagger_\om(0)
+\frac{i\om}{\sqrt{2}l_0}\int_{-\infty}^t dt'\,
e^{i\om(t-t')}\hat{\xi}(t'),
\label{T1.13}\eea
assuming that the field $\hat{\xi}(t)$ is absent at $t=-\infty$, and that the solution is retarded. The operators  $\hat{b}^\dagger_\om(0)$ and $\hat{b}_\om(0)$  are the initial operators fulfilling the commutator relation \Ref{T1.7}.

These operators allow to write down the coordinate operator for the bath oscillators also in Heisenberg representation, and from \Ref{T1.8} we get
\be \hat{q}_\om(t)=\hat{q}^{\rm h}_\om(t)+
\int_{-\infty}^t dt'\,\om\sin(\om(t-t')) \hat{\xi}(t')
\label{T1.14}\ee
with the homogeneous solution
\be \hat{q}^{\rm h}_\om(t)=\frac{l_0}{\sqrt{2}}
\left(e^{-i\om t}\hat{b}_\om(0)+e^{i\om t}\hat{b}^\dagger_\om(0)\right).
\label{T1.15}\ee
Now we insert the solution \Ref{T1.13}  into the right side of eq. \Ref{T1.12} for the oscillator and using \Ref{T1.15} we get
\be  m(\pa_t^2+\Om^2)\hat{\xi}(t)=
    F_{\rm L}(t)-\hat{\Gamma}\hat{\xi}(t),
\label{T1.16}\ee
where
\be F_{\rm L}(t)=\int_0^\infty d\om\,\mu\om^2
    \hat{q}^{\rm h}_\om(t)
\label{T1.17}\ee
is the well known Langevin force. With \Ref{T1.15} inserted, it can also be written in the form
\be F_{\rm L}(t)=\int_0^\infty d\om\,\frac{\mu\om^2l_0}{\sqrt{2}}
    \left(e^{-i\om t}\hat{b}_\om(0)+e^{i\om t}\hat{b}^\dagger_\om(0)\right).
\label{T1.18}\ee
The second term in the right side of \Ref{T1.16} is
\be   -\hat{\Gamma}\hat{\xi}(t) \equiv
    \int_0^\infty d\om\,\mu\om^2\left[\int_{-\infty}^t dt'\,\om\sin(\om(t-t'))\hat{\xi}(t')-\hat{\xi}(t)\right].
\label{T1.19}\ee
Here, $\hat{\Gamma}$ denotes a linear, positive operator as discussed in \cite{ford88-37-4419}. One may proceed using its general properties, but we found it much more instructive to restrict to the case of linear damping by parameterizing the mass density $\mu$ of the bath in the form
\be \mu=\frac{2\ga m}{\pi\om^2}\frac{1}{1+(\delta \om)^2},
\label{T1.20}\ee
where $\ga$ is the dissipation parameter and $\delta$ is a regularization parameter with $\delta\to0$ at the end.
Using the parametrization \Ref{T1.20}, we do the following transformations with expression
\be A\equiv  \int_0^\infty d\om\,\mu\om^2\int_{-\infty}^t dt'\,\om\sin(\om(t-t'))\hat{\xi}(t').
\label{T1.21}\ee
First, we integrate by parts  in in the integral over $t'$,
\be A= \int_0^\infty d\om\,\mu\om^2\hat{\xi}(t)
-\int_0^\infty d\om\,\mu\om^2\int_{0}^\infty dt'\,\cos(\om t') {\dot{\hxi}}(t-t'),
\label{T1.22}\ee
where we also made the substitution $t'\to t-t'$. We mention that the integrations in \Ref{T1.21} and \Ref{T1.22} are convergent for $\delta>0$. Now, in the second term in \Ref{T1.22} we make the substitutions $\om\to\om/\delta$ and $t'\to t' \delta$, and after that tend $\delta\to 0$. Using
\be \int_0^\infty d\om\,\frac{2\ga m}{\pi}\frac{1}{1+\om^2}
\int_0^\infty dt'\,\cos(\om t') {\dot{\hxi}}(t-\delta t')
\raisebox{-4pt}{$\to\atop\delta\to0$} \ga m  {\dot{\hxi}}(t),
\label{T1.23}\ee
we arrive at
\be A=\int_0^\infty d\om\,\mu\om^2\hat{\xi}(t)-\ga m {\dot{\hxi}}(t)
\label{T1.24}\ee
for $\delta\to 0$. Returning to \Ref{T1.19} and using \Ref{T1.21} we get for the linear damping \Ref{T1.20}
\be  -\hat{\Gamma}\hat{\xi}(t)=-\ga m {\dot{\hxi}}(t).
\label{T1.25}\ee
Taking this term to the left side of eq. \Ref{T1.16}, we get the equation
\be  m(\pa_t^2+\ga\pa_t+\Om^2)\hat{\xi}(t)=
    F_{\rm L}(t)
\label{T1.26}\ee
for a harmonic oscillator $\hat{\xi}(t)$, which is damped by $\ga$ and exited by the Langevin force $F_{\rm L}(t)$. This is just the well known approach of a heat bath coupled to an oscillator. The bath provides dissipation $\ga$ and at once, in equilibrium, it provides excitations that keep the oscillator in a steady state.

Eq. \Ref{T1.26} can be solved most simply using a Fourier transform. We use here  and below, for all time dependent quantities, the conventions
\be \xi(t)=\int_{-\infty}^\infty\frac{d\om}
    {2\pi}e^{i\om t}\txi_\om, \quad \txi_\om=\int_{-\infty}^\infty dt\,
    e^{-i\om t}\xi(t).
\label{T1.27}\ee
eq. \Ref{T1.26} turns into
\be mN(\om)\txi_\om=\tilde{F}_\om
\label{T1.28}\ee
with
\be N(\om) \equiv -\om^2+i\ga\om+\Om^2
\label{T1.29}\ee
and
\be \tilde{F}_\om= 2\pi\sqrt{\frac{\ga m}{\pi}\hbar \om}
    \left(\Theta(-\om)\hat{b}_{-\om}(0)
            +\Theta(\om)\hat{b}^\dagger_{\om}(0)\right)
\label{T1.30}\ee
is the Fourier transformed of the Langevin force \Ref{T1.18}.

Now the solution of eq. \Ref{T1.28} is
\be \txi_\om=\txi_\om^{\rm h}+\frac{1}{m N(\om)}\tilde{F}_\om,
\label{T1.31}\ee
where $\txi_\om^{\rm h}$ is the homogeneous solution. Transforming back to time, we get
\be \hat{\xi}(t)= \hat{\xi}^{\rm h}(t)
\label{T1.32}+
    \int_0^\infty d\om\,\sqrt{\frac{\ga}{\pi m}\hbar \om}
    \left(\frac{e^{-i\om t}}{N(-\om)}\hat{b}_\om
    +\frac{e^{i\om t}}{N(\om)}\hat{b}^\dagger_\om\right),
    \ee
with the homogeneous solution,
\be \hat{\xi}^{\rm h}(t)=\frac{1}{\sqrt{2\om_1}}
 \left({e^{-i\om_1 t}}\hat{a}
    +{e^{i\om_1 t}}\hat{a}^\dagger \right)e^{-(\ga/2) t},
\label{T1.32a}\ee
with $\om_1=\sqrt{\Om^2-(\ga/2)^2}$ and $[\hat{a},\hat{a}^\dagger]=1$.
Here, the physics is obvious: the homogeneous solution dies out with time due to the damping, and the inhomogeneous solution   of the oscillator is supported by the Langevin force.

\subsection{\label{secT1.2}Thermal averages and free energy}
In a thermodynamic system, the basic potentials are the (internal) energy,
\be E(S,V),~~\mbox{with}~T=\frac{\pa E}{\pa S},
\label{T1.33}\ee
and the free energy,
%.
\be F(T,V),~~\mbox{with}~S=-\frac{\pa F}{\pa T}.
\label{T1.34}\ee
These are related by a Legendre transform, $F=E-TS$.
In statistical quantum mechanics, for a canonical ensemble, one defines the density operator,
\be \hat{\rho}=\frac{1}{Z}e^{-\beta \hat{H}},
\label{T1.35}\ee
with the inverse temperature $\beta=1/T$ and
\be Z={\rm Tr} \hat{\rho}.
\label{T1.36}\ee
Within this approach, the energy is the average,
\be E=\langle \hat{H}\rangle={\rm Tr} \hat{H}\hat{\rho},
\label{T1.37}\ee
and the free energy is given by
\be F=-T\ln(Z),
\label{T1.38}\ee
and the relation
\be
E=\frac{\pa}{\pa \beta}(\beta F)
\label{T1.39}\ee
holds in agreement with \Ref{T1.34} and \Ref{T1.38}.

In the following, for the systems considered in this paper, we are going to calculate the energies as averages like \Ref{T1.37} and the free energy via relations like \Ref{T1.39}

The basic thermal averages appear from the bath oscillators
obeying \Ref{T1.7} with the Hamilton operator \Ref{T1.10}
and the well known formulas
\be Z=\frac{e^{-\beta\hbar\om/2}}{1-e^{-\beta\hbar\om}},
\label{T1.40}\ee
and
\be \langle \hat{b}_\om \hat{b}^\dagger_{\om'}\rangle =
\delta(\om-\om')\left(\frac{1}{e^{\beta\hbar\om}-1}+1\right),
\quad\langle \hat{b}^\dagger_\om \hat{b}_{\om'}\rangle =
\delta(\om-\om') \frac{1}{e^{\beta\hbar\om}-1},
\label{T1.41}\ee%
hold, giving together
\be \langle \hat{b}_\om \hat{b}^\dagger_{\om'}
+ \hat{b}^\dagger_\om \hat{b}_{\om'}\rangle  =
\delta(\om-\om'){\cal N}_T(\om)
\label{T1.42}\ee
with
\be {\cal N}_T(\om) \equiv \frac{2}{e^{\beta\hbar\om}-1}+1=\coth\frac{\beta\hbar\om}{2}.
\label{T1.43}\ee
For a simple bosonic harmonic oscillator with
\be \hat{H}_{\rm bath}= \frac{\hbar\om}{2}(\hat{b}^\dagger_\om\hat{b}_\om+\hat{b}_\om\hat{b}^\dagger_\om).
\label{T1.44}\ee
we remind  the energy
\be E=\frac{\hbar\om}{2}{\cal N}_T(\om),
\label{T1.45a}\ee
and the free energy
\be F=\frac{\hbar\om}{2}+T\ln\left(1-e^{-\beta\hbar\om}\right).
\label{T1.45}\ee
\subsection{\label{secT1.3}Energy flows}
Having with eqs. \Ref{T1.14} and \Ref{T1.32} the time dependent solutions of the Heisenberg equations of motion \Ref{T1.11} and \Ref{T1.12}, we can investigate the energies \Ref{T1.3a}-\Ref{T1.3c}. First, we look at eq. \Ref{T1.6} and multiply by $\dot{\hq}_\om(t)$, such that the equation can be written in the form
\be \frac{d}{d t}\frac{\mu}{2}\left(\dot{\hq}_\om(t)^2+\om^2 \hq_\om(t)^2\right)  =
    \mu\om^2\dot{\hq}_\om(t)\hxi(t).
\label{T1.46}\ee
Integrating over $\om$, using \Ref{T1.3b} and \Ref{T1.4}, we come to
\be \frac{d}{d t}H_{\rm bath}  =
    \int_0^\infty d\om\,\mu\om^2\dot{\hq}_\om(t)\hxi(t).
\label{T1.47}\ee
In a similar way, we get from eq. \Ref{T1.5}, multiplying by $\dot{\hxi}(t)$, and using \Ref{T1.3a} and \Ref{T1.4},
\be \frac{d}{d t}H_{\rm osc}  =
    \int_0^\infty d\om\,\mu\om^2 ( {\hq}_\om(t)-\hxi(t))\dot{\hxi}(t).
\label{T1.48}\ee
Next, we insert the solutions \Ref{T1.14} into \Ref{T1.47},
\be  \frac{d}{d t}H_{\rm bath}  =
    \int_0^\infty d\om\, \mu \om
    \left[ \dot{\hq}^{\rm h}_\om(t)
     \label{T1.49}  +\int_{-\infty}^{t}dt'\,\om^2\cos(\om(t-t'))\hat{\xi}(t')
    \right]\hxi(t).
\ee
We use \Ref{T1.17} and integrate by parts,
\be \frac{d}{d t}H_{\rm bath}  = \dot{F}_{\rm L}(t)\hxi(t)
 +\int_0^\infty d\om\,\mu\om^2
    \int_{-\infty}^{t}dt'\,\om\sin(\om(t-t')) {\dot{\hxi}}(t')
    \hxi(t).
\label{T1.50}\ee
Further, using \Ref{T1.21}, and acting similar as in deriving \Ref{T1.24}, we get
\be \frac{d}{d t}H_{\rm bath}  = (\dot{F}_{\rm L}(t)-\ga m\ddot{\hxi}(t))\hxi(t)
    +\int_0^\infty d\om\,\mu\om^2
    {\dot{\hxi}}(t)    \hxi(t).
\label{T1.51}\ee
Next, we insert the solution \Ref{T1.14} into \Ref{T1.48}, use \Ref{T1.21} and get for the oscillator
\be \frac{d}{d t}H_{\rm osc}  =
    \int_0^\infty d\om\,\mu\om^2 \hat{q}^{\rm h}_\om(t)\dot{\hxi}(t)
    +A \dot{\hxi}(t)
    - \int_0^\infty d\om\,\mu\om^2 {\hxi}(t)\dot{\hxi}(t),
\label{T1.52}\ee
which with \Ref{T1.24} gives
\be \frac{d}{d t}H_{\rm osc}  =
     ({F}_{\rm L}(t)-\ga m\dot{\hxi}(t))\dot{\hxi}(t).
\label{T1.53}\ee
Finally, we insert \Ref{T1.14} into \Ref{T1.3c} and using \Ref{T1.21}
we have
\be  H_{\rm int}  = -\int_0^\infty d\om\,\mu\om^2
\hat{q}_\om^{\rm h}\hxi(t)-A\hxi(t)
+\frac12 \int_0^\infty d\om\,\mu\om^2\hxi(t)^2,
\label{T1.54}\ee
which with \Ref{T1.17} and \Ref{T1.24} turns into
\be  H_{\rm int}  =-(F_{\rm L}(t)-m\ga\dot{\hxi}(t))\hxi(t)-\frac12 \int_0^\infty d\om\,\mu\om^2\hxi(t)^2.
\label{T1.55}\ee
Now, adding \Ref{T1.51}, \Ref{T1.53} and \Ref{T1.55}, we see that the total energy is time conserved,
\be \frac{d}{d t}H=\frac{d}{d t}H_{\rm bath}
    +\frac{d}{d t}H_{\rm osc}+\frac{d}{d t}H_{\rm int} =0.
\label{T1.56}\ee
Since the individual contributions in \Ref{T1.56} are non zero, we have  energy flows between them compensating each other at any time.
\subsection{\label{secT1.4}Internal and free energies of the oscillator}
We use the definitions given in subsection \ref{secT1.1} and calculate the thermal averages of the Hamiltonians \Ref{T1.3a}-\Ref{T1.3c}. Since we are interested in equilibrium, we insert only the inhomogeneous part of the solution \Ref{T1.32},
\be \hat{\xi}(t)=
    \int_0^\infty d\om\,\sqrt{\frac{\ga}{\pi m}\hbar \om}
    \left(\frac{e^{-i\om t}}{N(-\om)}\hat{b}_\om
    +\frac{e^{i\om t}}{N(\om)}\hat{b}^\dagger_\om\right).
    \label{T1.57}\ee
First of all, we use the results of the preceding subsection to mention that we have with \Ref{T1.56}
\be  \frac{d}{d t}H_{\rm bath}
    +\frac{d}{d t}H_{\rm int} =-\frac{d}{d t}H_{\rm osc}.
\label{T1.58}\ee
Below we will see that $\frac{d}{d t}\langle H_{\rm osc}\rangle=0$
holds. Thus the energy $\langle H_{\rm bath}+H_{\rm int}\rangle={\rm const}$, i.e., it does not depend on time and can be determined at $t\to-\infty$, where according to our initial conditions we have only the energy of the bath. As a consequence, the internal energy, we are interested in, is given by
\be  E=\langle H_{\rm osc}\rangle,
\label{T1.59}\ee
with $H_{\rm osc}$ given by \Ref{T1.3a} and the averages by \Ref{T1.42}.

We insert solution \Ref{T1.57} into \Ref{T1.59} and get a double integration in $\om$, which by means of the averages \Ref{T1.42} gets reduced to a single integration. This way we get
\be  E=\langle H_{\rm osc}\rangle
    = {\ga}\int_0^\infty d\om\, \frac{\hbar\om}{2}{\cal N}_T(\om)
    \frac{\om^2+\Om^2}{ |  N(\om) | ^2},
\label{T1.59a}\ee
where $N(\om)=N(-\om)^*$ was used. eq. \Ref{T1.59a} can be simplified using the identity
\be  \ga\frac{\om^2+\Om^2}{ |  N(\om) | ^2}
    =\frac{\pa}{\pa\om}\frac{1}{2i}\ln\frac{N(\om)}{N(-\om)}.
\label{T1.60}\ee
The structure of this expression makes it meaningful to define a phase,
\be \delta(\om) = \frac{1}{2i}\ln\frac{N(\om)}{N(-\om)},
\label{T1.60a}\ee
which allows one to write the energy in the form
\be  E
    = \int_0^\infty \frac {d\om}{2\pi}\, {\hbar\om}
    {\cal N}_T(\om)\frac{\pa}{\pa\om}\delta(\om) ,
\label{T1.61}\ee
which is the basic average for the energy to be used below.
It is to be mentioned that the average \Ref{T1.59a}, and all alike,  are time independent.

The free energy of the oscillator and the bath can be obtained from \Ref{T1.61} using \Ref{T1.39},
\be  F  =
T\int_0^\infty \frac{d\om}{\pi}\,
      %  \frac{\pa}{\pa\om}\left(\frac{1}{2i}\ln\frac{N(\om)}{N(-\om)} \right)
\left(\frac{\beta\hbar\om}{2}+\ln\left(1-e^{-\beta\hbar\om}\right)\right)\frac{\pa}{\pa\om}\delta(\om).
\label{T1.62}\ee
Integrating by parts, one comes to a different representation,
\be F=-\hbar\int_0^\infty \frac{d\om}{2\pi}\,
    {\cal N}_T(\om)  \delta(\om),
\label{T1.63}\ee
and looking at \Ref{T1.43} one may separate zero temperature and temperature dependent parts,
\be F=F_0+\Delta_TF,
\label{T1.63a}\ee
with
\be F_0 =-\hbar\int_0^\infty \frac{d\om}{2\pi}\, \delta(\om),
   \quad
\Delta_TF =
    -\hbar\int_0^\infty \frac{d\om}{\pi}\,
\frac{1}{e^{\beta\hbar\om}-1}  \, \delta(\om).
\label{T1.63b}\ee
It should be mentioned, that internal and free energies,  \Ref{T1.61}, \Ref{T1.62} or \Ref{T1.63},  are expressed in terms of real frequencies even in case of dissipation, $\ga>0$.
In general, such representations are not new. For instance, eq. \Ref{T1.59a} is similar to eq. (64) in \cite{rosa10-81-033812} and eq. \Ref{T1.61} is similar to eq. (5) in \cite{ford85-55-2273}, where it was called 'remarkable formula". What I called 'phase', $\delta(\om)$, \Ref{T1.60a}, is expressed as imaginary part of a Green's function in the mentioned papers.
\subsection{\label{secT1.5}Special cases}
In this subsection we consider some special cases. First, we switch off dissipation, $\ga\to0$. For this we return to eq. \Ref{T1.60} and note
\be \frac{\ga}{ |  N(\om) | ^2}=\frac{1}{2i\om}\left(
    \frac{1}{N(-\om)}- \frac{1}{N(\om)}\right)
    \raisebox{-4pt}{$=\atop\ga\to0$}\frac{\pi}{\om}\delta(\om^2-\Om^2),
\label{T1.64}\ee
where we used the Sokhotski-Plemelj theorem, such that \Ref{T1.60} becomes
\be \frac{\pa}{\pa\om}\left(\frac{1}{2i}\ln\frac{N(\om)}{N(-\om)}
    \right)
    \raisebox{-4pt}{$=\atop\ga\to0$}\pi\delta(\om-\Om).
\label{T1.65}\ee
Here we also accounted for $\om\ge0$ in the integration in \Ref{T1.62}.

This way, without dissipation, the free energy becomes
\be F_{|_{\ga=0}}=\frac{\hbar\Om}{2}+T\ln\left(1-e^{-\beta\hbar\Om}\right),
\label{T1.66}\ee
in agreement with \Ref{T1.45} for a single oscillator in a thermal state.

Next we consider $T\to0$ and use representation \Ref{T1.62} for the temperature dependent part. Here we can expand
\be  \frac{1}{2i}\ln\frac{N(\om)}{N(-\om)}
=\ga\frac{\om}{\Om^2}+\dots \,,
\label{T1.67}\ee
for small $\om$, which gives in \Ref{T1.63b} the first term in the expansion for $T\to0$,
\be \Delta_T F       \raisebox{-4pt}{$=\atop T\to0$}
    -\frac{T^2\ga}{\hbar\Om^2}\frac{\zeta(2)}{\pi}+\dots \,,,
\label{T1.68}\ee
where $\zeta(2)=\pi^2/6$ is the Riemann Zeta function.
The  same can be done for $\Om=0$. Using
\be  \frac{1}{2i}\ln\frac{N(\om)}{N(-\om)}_{|_{\Om=0}}
      =\frac{\om}{\ga}+\dots \,
\label{T1.69}\ee
we get
\be \Delta_T F_{|_{\Om=0}} \raisebox{-4pt}{$=\atop T\to0$}
    -\frac{T^2}{\hbar\ga}\frac{\zeta(2)}{\pi}+\dots \,.
\label{T1.70}\ee
In both cases, we get for the entropy \Ref{T1.34} for $T\to0$
\be
    S=\frac{2T\ga}{\hbar\Om^2}\frac{\zeta(2)}{\pi}+\dots \,,\quad
    S_{|_{\Om=0}} =    \frac{2T}{\hbar\ga}\frac{\zeta(2)}{\pi}+\dots \,.
\label{T1.71}\ee
It should be mentioned that it is not meaningful to consider here the limit $\ga\to0$ since it returns one to a single oscillator as mentioned on eq. \Ref{T1.66}.

\section{\label{T2}Two oscillators interacting through a field}
In this section we consider a system consisting of two oscillators interacting through a field. The Lagrangian is
\bea {\cal L}&=&\int dx\, \frac12\left(\dot{\phi}(t,x)^2-{\phi'(t,x)}^2\right)
 \label{T2.1}\\&&   +\sum_{i=1}^2\frac{m}{2}\left(\dxi_i(t)^2-\Om^2\xi_i(t)^2\right)
    -e\sum_{i=1}^2\xi_i(t)\phi(t,a_i),
\nn\eea
where $\phi(t,x)$ is the field, $\dot{\phi}$ and $\phi'$ are its temporal and spatial derivatives. The oscillators are described by their displacements $\xi_i(t)$, intrinsic frequency $\Om$ and location $a_i$. The interaction strength is denoted by $e$. The equations of motion following from \Ref{T2.1} are
\bea    \left(\pa_t^2-\pa_x^2\right)\phi(t,x) &=&
    -e\sum_{i=1}^2 \xi_i(t)\delta(x-a_i), \nn \\
    m\left(\pa_t^2+\Om^2\right)\xi_i(t) &=&
    -e \phi(t,a_i),\quad (i=1,2).
\label{T2.2}\eea
This system resembles the plasma model in electrodynamics, where the dipoles (oscillators) obey an equation of motion similar to that of $\xi_i(t)$ in \Ref{T2.2}.

In general, the system \Ref{T2.2} can be solved in two ways. In the \underline{1$^{\rm st}$ way}, one solves the second equation \Ref{T2.2} for $\xi_i(t)$ and inserts the result into the first equation, which gives an effective equation for the field $\phi(t,x)$. In the \underline{2$^{\rm nd}$ way}, one solves first the equation for $\phi(t,x)$  and inserts the result into the second equation which gives an effective equation for $\xi_i(t)$.

The {1$^{\rm st}$ way} corresponds to the conventional approach to the Casimir-Polder forces, while the {2$^{\rm nd}$ way} corresponds to some of atomistic approach. As a kind of peculiarity of the {2$^{\rm nd}$ way}, it turns out that it requires, for consistency reasons,  to put the field into a finite box as we will see below.

Before going into detail of the two ways, we rewrite the equations \Ref{T2.2} in Fourier representation using the conventions \Ref{T1.27},
\bea    \left(-\om^2-\pa_x^2\right)\tphi_\om(x) &=&
    -e\sum_{i=1}^2 \txi_{i\om}\delta(x-a_i), \label{T2.3} \\
    m\left(-\om^2+\Om^2\right)\txi_{i\om} &=&
    -e \tphi_\om(a_i),\quad (i=1,2).
\label{T2.4}\eea
These are the formulas, which we will use in the following.
\subsection{\label{secT2.1}{1$^{\rm st}$ way}}
We solve equation \Ref{T2.4}. Its inhomogeneous solution is
\be \txi_{i\om}=-\frac{e}{m \kern 0.05em  N(\om)}\,\tphi_\om(a_i),
\label{T2.5}\ee
with %
\be N(\om)=-\om^2+\Om^2.
\label{T2.6}\ee
It is to be mentioned, that its homogenous solution would give a contribution to the energy which does not depend on the separation $b$ between the oscillators and which is irrelevant for what we are interested in.
Insertion of \Ref{T2.6} into \Ref{T2.3} delivers
\be \left(-\om^2-\pa_x^2-\frac{e^2}{
            m\kern 0.05em N(\om)
}\sum_{i=1}^2\delta(x-a_i)\right)
    \tphi_\om(x)=0,
\label{T2.7}\ee
which is the effective equation for the field accounting for the interaction with the oscillators. Frequently, one defines with
\be \al(\om)=\frac{e^2}{
            m\kern 0.05em N(\om)
}
\label{T2.8}\ee
a dynamic polarizability of the oscillators, which allows to write equation \Ref{T2.7} also in the form
\be \left(-\om^2-\pa_x^2-\al(\om)\sum_{i=1}^2\delta(x-a_i)\right)
    \tphi_\om(x)=0.
\label{T2.9}\ee
This equation is the same as in the text book examples of a Schr{\"o}dinger equation with two delta function potentials.
The solution to eq. \Ref{T2.7} can be represented by
\be \tphi_\om(x) =
    e^{i\om x}+
    \frac{e^2}{m}\sum_{r,s=1}^2 G_0^{(\infty)}(x-a_r)
    \Phi^{-1}_{rs}(\om)  \,e^{i\om a_s},
\label{T2.10}\ee
where
\be G_0^{(\infty)}(x)=\frac{e^{i\om |  x | }}{-2i\om}
\label{T2.11}\ee
is a 'free' Green's function obeying
\be \left(-\om^2-\pa_x^2\right)G_0^{(\infty)}(x)=\delta(x)
\label{T2.12}\ee
for $x\in (-\infty,\infty)$, i.e., on the whole $x$-axis and $\Im \om>0$ is assumed. We introduce the notation
\be \Phi_{rs}(\om) = N(\om)\delta_{rs}-\frac{e^2}{m}\,G_0^{(\infty)}(a_r-a_s),
\label{T2.13}\ee
and inserting solution \Ref{T2.10} into equation \Ref{T2.7}, and using \Ref{T2.12},  it can be checked easily that $\Phi^{-1}_{rs}(\om)$ is the inverse matrix to $\Phi_{rs}(\om) $, \Ref{T2.13}.

Solution \Ref{T2.10} is a scattering solution. It has asymptotic expansions
\bea \tphi_\om(x) &\raisebox{-4pt}{$\sim\atop x\to -\infty$}
                    &       e^{i\om x}+r(\om)\,e^{-i\om x},
    \nn\\
    \tphi_\om(x) &\raisebox{-4pt}{$\sim\atop x\to  \infty$}
                    &    t(\om) \,  e^{i\om x},
\label{T2.14}\eea
where $r(\om)$ and $t(\om)$ are the reflection and transmission coefficients.
Using these formulas, the transmission coefficient
\be t(\om) = 1+ \frac{e^2}{m}\sum_{r,s=1}^2 \frac{e^{i\om(a_s-a_r)}}{-2i\om}
    \Phi^{-1}_{rs}(\om)
\label{T2.15}\ee
can be read off.
Further, using \Ref{T2.13}, after some algebra, \Ref{T2.15} can be rewritten in the form
\be t(\om) =   \frac{1}{\left(1+\frac{\al(\om)}{2i\om}\right)^2
                    -\left( \frac{\al(\om)}{2i\om}
                    e^{i\om b}\right)^2  },
\label{T2.16}\ee
with $\al(\om)$ given by \Ref{T2.8} and $b=a_2-a_1$ is the separation between the oscillators.

With the transmission coefficient \Ref{T2.16} we have the complete information we need for the vacuum energy of the field $\phi(t,x)$ obeying eq. \Ref{T2.9}. The procedure, of how to obtain the vacuum energy in the case of a given transmission coefficient,  is described in many places, we follow here the paper \cite{bord95-28-755} (eqs. (14), (15), with the notation $s_{11}(k)\to t(\om)$). Defining the scattering phase by
\be \delta(\om)=\frac{1}{2i}\ln\frac{t(\om)}{t(-\om)},
\label{T2.17}\ee
for the vacuum energy the formula
\be E_0=\frac{\hbar}{2}\int_0^\infty\frac{d\om}{\pi}\,
\om^{1-2\ep}\pa_\om\delta(\om)
\label{T2.18}\ee
holds, where $\ep>0$ is an ultraviolet regularization, $\ep\to0$ at the end.

It should be mentioned, that eq. \Ref{T2.9}, besides the scattering solution \Ref{T2.10}, also has bound solutions, which are given by the poles of $t(\om)$, \Ref{T2.16} on the imaginary axis, say at $\om=i\kappa_b$. It can be shown that the equation
\be t(i\kappa)^{-1}=0
\label{T2.18a}\ee
has at least one solution for all values of the parameters entering \Ref{T2.16}. These   bound   solutions have negative energy and will be occupied by $\phi$-particles in the ground state of the system forming a kind of Dirac sea. Therefore we do not need to include them into the vacuum energy \Ref{T2.18}.

It is convenient to turn in \Ref{T2.18} the integration towards imaginary frequencies. For this we use \Ref{T2.17} and write $\delta(\om)$ as a difference of two logarithms. In the first one we turn the integration contour upwards, $\om\to i\xi$, and in the second, downwards, $\om\to -i\xi$. These directions were predefined by the choice $\Im \om>0$ in \Ref{T2.11} and correspond to a Wick rotation. Following the steps described in \cite{bord95-28-755}, we arrive at
\be E_0=\frac{\hbar}{2}\cos(\pi \ep)\int_0^\infty\frac{d\xi}{\pi}\,
    \xi\pa_\xi  \ln  |  t(i\xi) |  +
    \frac{\sin(\pi \ep)}{2\pi}\kappa_b^{1-2\ep},
\label{T2.19}\ee
where $\kappa_b$ is a solution of equation \Ref{T2.18a}.

Now it is possible to rewrite the energy \Ref{T2.19}, dropping a contribution which does not depend on the sepa\-ration $b$ between the oscillators, by rewriting $t(\om)$ in the form
%%
%\be t(i\xi) = \left[1-\left(\frac{\al(i\xi)}{2\xi-\al(i\xi)}e^{-\xi b}\right)^2\right]^{-1}.
%\label{T2.20}\ee
%%
%
\be t(\om) = \left[1-\left(\frac{\al(\om)}{2i\om+\al(\om)}e^{i \om  b}\right)^2\right]^{-1}.
\label{T2.20}\ee
After that, because of the decrease of $ \ln  |  t(i\xi) | $ with $t(\om)$ given by eq. \Ref{T2.20} for $\xi\to\infty$,  we may remove the regularization and put $\ep=0$. In addition, we integrate by parts and represent the vacuum energy in the form
\be E_0=\frac{\hbar}{2}\int_0^\infty
\frac{d\xi}{\pi}\,
    \ln  |  t(i\xi)^{-1} | ,
\label{T2.21}\ee
with, explicitly written using \Ref{T2.8},
\be t(i\xi)^{-1}=1-\left(\frac{e^2/m}{2\xi(\xi^2+\Om^2)-e^2/m}\,
e^{-\xi b}\right)^2.
\label{T2.22}\ee
Equations \Ref{T2.21} and \Ref{T2.22} are a way of writing for the vacuum energy which one knows from the Casimir-Polder force between to oscillators. We mention also the special case
\be E_0=-\frac{\pi^2\hbar}{24b},
\label{T2.23}\ee
which one obtains from \Ref{T2.21} for $b\to\infty$ or $e^2\to\infty$, when the problem turns into that on an interval of length $b$ with Dirichlet boundary conditions.

\subsection{\label{secT2.2}{2$^{\rm nd}$ way}}
To start with, we consider the problem on the whole axis. The solution of eq. \Ref{T2.3} is given by %
\be \tphi_\om(x)=-e\sum_{i=1}^2G_0^{(\infty)}(x-a_i)\txi_{i\om}
\label{T2.24}\ee
with $G_0^{(\infty)}(x)$ given by eq. \Ref{T2.11}. Insertion into  \Ref{T2.2} gives
\be \sum_{j=1}^2\left(\left(-\om^2+\Om^2\right)\delta_{ij}
-\frac{e^2}{m} G_0^{(\infty)}(a_i-a_j)
\right)
\txi_{j\om}=0,
\label{T2.25}\ee
which is a set of algebraic equations. In fact, the factor in front of $\txi_{j\om}$ in \Ref{T2.25} is just $\Phi_{ij}(\om)$, eq. \Ref{T2.13}. Thus, the eigenfrequencies of \Ref{T2.25} are given by the zeros $\om_s$ of the determinant
\be \det \Phi_{ij}(\om)=0 \quad \Rightarrow
\quad \om=\om_s,
\label{T2.26}\ee
and the corresponding vacuum energy is simply
\be E_0=\frac{\hbar}{2}\sum_s\om_s^{1-2\ep},
\label{T2.27}\ee
where, again, $\ep$ is an ultraviolet regularization.

However, equation \Ref{T2.26} with $G_0^{(\infty)}(x)$, \Ref{T2.11},  does not have real zeros, as can be seen easily by inspection. Thus, \Ref{T2.27} does not provide a real vacuum energy. This circumstance was  already mentioned in \cite{renn71-53-193}. There it was pointed out that this happens since the oscillators, interacting through the field, will radiate if placed in an infinite volume. As a way out, one may put the whole system in a finite box. In that case the eigenfrequencies $\om_s$, and with them the vacuum energy \Ref{T2.27}, will be real. Afterwards, one may take the size of the box to infinity and look on the separation dependence of the vacuum energy. We will carry out this procedure in the following.
We realize the 'finite box' as interval $x\in [-\frac{L}{2},\frac{L}{2} ]$ and place the oscillators at $x=a_{1,2}$ with
\be a_{1,2}=\mp\frac{b}{2},
\label{T2.28}\ee
where $b=a_2-a_1$ is the separation between the oscillators. On the boundaries we demand the field $\phi(t,x)$ to fulfil Dirichlet boundary conditions,
\be \phi(t,\pm {L}/{2})=0.
\label{T2.29}\ee
We need  the Green's function $G_0(x)$, fulfilling the equation
\be \left(-\om^2-\pa_x^2\right)
    G_0 (x,x')=\delta(x-x'),
\label{T2.30}\ee
and the boundary conditions \Ref{T2.29}. It can be constructed out of the mode functions
\be \varphi_n(x)=\sqrt{\frac{2}{L}}\sin\left(k_n\left({x}-\frac{L}{2}\right)\right),
\label{T2.31}\ee
with $k_n=\frac{\pi n}{L}$.
These form a basis, fulfilling
\be \int_{-L/2}^{L/2} dx\, \varphi_{n}(x)\varphi_{n'}(x)=\delta_{nn'},
\label{T2.32}\ee
and the Green's function is
\be G_0 (x,x')=\sum_{n=1}^\infty \frac{\varphi_n(x)\varphi_n(x')}{-\om^2+k_n^2}.
\label{T2.33}\ee
In general, the sum in \Ref{T2.33} can be carried out. For $|x|<L$ and $|x'|<L$, we get
\be G_0 (x,x') \label{T2.33x}
= \frac{e^{i\om |x-x'|}-e^{-i\om |x+x'|}}{-2i\om}
-\frac{\sin(\om x)\sin(\om x')}{\om\sin(\om L)}\,e^{i\om L}.
\ee
But we don't need the corresponding formulas  in general form and mention the two special cases occurring in \Ref{T2.36}. Due to the symmetric positions \Ref{T2.28} of the oscillators, these cases are
\bea G_0\left(\pm\frac{b}{2},\pm\frac{b}{2}\right) \equiv G_1(\om)&=&
\frac{\cos(\om b)-\cos(\om L)}{2\om\sin(\om L)},
\nn\\
G_0\left(\mp\frac{b}{2},\pm\frac{b}{2}\right) \equiv G_2(\om)&=&
\frac{1-\cos(\om (L-|b|))}{2\om\sin(\om L)},~~~~
\nn\\\label{T2.33a}\eea
where we introduced with $G_{1,2}(\om)$ special notations for these quantities. Note, the  formulas \Ref{T2.33a} are valid for $|b|<L$.

Using the Green's function \Ref{T2.33}, we get the solution of  equation \Ref{T2.3}, obeying the boundary conditions \Ref{T2.29}, in the form
\be \tphi_\om(x)=-e\sum_{i=1}^2G_0 (x,a_i)\txi_{i\om}.
\label{T2.34}\ee
We insert this solution into eq. \Ref{T2.4}, and get in place of \Ref{T2.25} the system of equations
\be \sum_{j=1}^2\left(\left(-\om^2+\Om^2\right)
\delta_{ij}
-\frac{e^2}{m} G_0(a_i,a_j)
\right)
\txi_{j\om}=0.
\label{T2.35}\ee
Again, the coefficients in front of $\txi_{j\om}$ can be written in the form \Ref{T2.13} with
\be \Phi_{rs}(\om) = N(\om)\delta_{rs}-\frac{e^2}{m}\,G_0(a_r,a_s).
\label{T2.36}\ee
The further procedure is straightforward. Define the frequencies $\om_s$ as  zeros,
\be \det \Phi_{ij}(\om)=0 \quad \Rightarrow
\quad \om=\om_s,
\label{T2.37}\ee
of the determinant of the system \Ref{T2.35},
one has a vacuum energy,
\be E_0=\frac{\hbar}{2}\sum_s\om_s^{1-2\ep},
\label{T2.38}\ee
like \Ref{T2.27}, but now the frequencies $\om_s$ are real, and, consequently, the energy $E_0$, \Ref{T2.38}, is real. It should be mentioned, that the realness of the $\om_s$ follows from the fact, that we have a hermite operator on a finite interval.

The next step is to consider in $E_0$, eq. \Ref{T2.38}, the limit $L\to\infty$. For this, we rewrite the sum over $s$ in \Ref{T2.38} as a contour integral, using \Ref{T2.37} as a mode generating function,
\be E_0=\frac{\hbar}{2}
    \int_\ga\frac{d\om}{2\pi i}\,
    \om^{1-2\ep}\pa_\om\ln\det\Phi_{ij}(\om),
\label{T2.39}\ee
where the path $\ga$ encircles the real positive frequency axis. The logarithm can be rewritten, using the fact that $\Phi_{ij}(\om)$, \Ref{T2.36}, is a $(2\times 2)$-matrix, as
\bea&& \ln\det\Phi_{ij}(\om)=
{\rm Tr}\ln \Phi_{ij}(\om)\label{T2.40}\\
&&= \ln\left[\left(N(\om)-\frac{e^2}{m}G_0(a_1,a_1)\right)^2
-\left(\frac{e^2}{m}G_0(a_1,a_2)\right)^2\right].
\nn\eea
In \Ref{T2.39} we divide the integration path $\ga$ into a part, $\ga_1$, in the upper half plane (with $\Im \om>0$) and a second path, $\ga_2$, in the lower half plane (with $\Im \om<0$), and consider $L\to\infty$ using \Ref{T2.33a}.
We get
\be \ln\det\Phi_{ij}(\om)
\raisebox{-4pt}{$\sim\atop L\to\infty$}
\left(N(\om)-\frac{e^2}{m}\frac{1}{-2i\om}\right)^2
-\left(\frac{e^2}{m}\frac{e^{i\om b}}{-2i\om}\right)^2
\label{T2.41}\ee
on $\ga_1$ and the complex conjugate on $\ga_2$. Now we finally turn the integration path towards the imaginary axis, $\om\to i\xi$ on $\ga_1$, and $\om\to -i\xi$ on $\ga_2$, and get from \Ref{T2.39}
\bea E_0&=&-\frac{\hbar}{2}\cos(\pi\ep)\int_0^\infty\frac{d\xi}{\pi}\,
\xi^{1-2\ep} \pa_\xi \ln
\label{T2.42}\\&&  \times
\left[
\left(N(i\xi)-\frac{e^2/m}{2\xi}\right)^2
%\right. \nn\\ && ~~~~~~~~~~~~~~~~~~~\left.
-\left(\frac{e^2/m}{2\xi}
e^{-\xi b}\right)^2
\right],
\nn\eea
after a calculation similar to that in the preceding subsection.
This equation is the final formula for the vacuum energy of the two dipoles, interaction through the field  $\phi(t,x)$, for $x$ on the whole axis. We remind the reader, that we were forced to put the field $\phi(t,x)$ first into a finite box,
$x\in [ -\frac{L}{2},\frac{L}{2} ]$. This way we got a real vacuum energy \Ref{T2.39}, which did depend on the size, $L$, of the box.

It should be mentioned, that for a sufficiently small box, $L<L_*$, with
\be   L_*=b+\frac{2m}{e^2}\Om^2,
\label{T2.43}\ee
there is no real solution of eq. \Ref{T2.18a} and the contribution from the bound solution in \Ref{T2.19} is absent. When increasing the box size $L$ beyond $L_*$, one or, in dependence on the parameters $\Om$ and $b$, two,  bound solutions appear. As already mentioned, we consider a ground state where these are occupied. As we will see in section \ref{T3}, in that case we have to take in \Ref{T2.42} the modulus of the square bracket.
%
%, and gave in the limit $L\to\infty$ the expression \Ref{T2.43}.

Finally, we compare \Ref{T2.42} with \Ref{T2.21}.  Using \Ref{T2.6}, it can be seen that taking out  from the logarithm in \Ref{T2.43} a contribution $\ln\left(N(i\xi)-\frac{e^2/m}{2\xi}\right)$, which is independent on the separation $b$, and dropping it, we can take $\ep=0$ in \Ref{T2.42}, because now the logarithm is decreasing for $\xi\to\infty$. Accounting for taking the modulus, and integrating by parts, we get from \Ref{T2.42} the same expression \Ref{T2.21} for the vacuum energy as in the preceding subsection.

%we come just to \Ref{T2.21} with \Ref{T2.22}.

This way, we have seen that the 1$^{\rm st}$ and the 2$^{\rm nd}$ ways, considered in this section, are equivalent in giving the same separation dependent part of the vacuum energy.
This equivalence was discussed also in \cite{renn67-1-317} in an atomistic derivation of the van der Waals interaction and it was in \cite{renn71-53-193} generalized to the retarded case.

\section{\label{T3}Two oscillators coupled to a heat bath and interacting through a field}
In this section, we consider the complete system consisting of two oscillators interacting with a heat bath each and with a field. Except for the heat bath, this is the same setup as in the preceding section. The interaction of one oscillator with a heat bath was considered in section \ref{T1}. In section \ref{T2} we were forced to work in a 'finite box', since otherwise we would not have real eigenfrequencies for the oscillators. In this section, there is strictly speaking no such motivation since in equilibrium all excitations are forced by the Langevin forces.
Nevertheless, also in this section we continue to consider the system in a finite box. As we will see below and in the next section, the transition to a large box is not trivial.
Also we mention, that on a formal level, as long as we work in terms of the Green's function $G_0(x,x')$, there is no difference in the formulas.

The Lagrangian for the complete system reads
\bea {\cal L} &=&\int dx\, \frac12\left(\dot{\phi}(t,x)^2-{\phi'(t,x)}^2\right)
 +\sum_{i=1}^2\frac{m}{2}\left(\dxi_i(t)^2-\Om^2\xi_i(t)^2\right)
\nn\\&&    -e\sum_{i=1}^2\xi_i(t)\phi(t,a_i)
       +\sum_{i=1}^2\int_0^\infty d\om\,\frac{\mu}{2}\left(\dot{q}_{i\om}(t)^2
-\om^2\left(q_{i\om}(t)-\xi_i(t)\right)^2\right),
\label{T3.1}  \eea
where $\phi(t,x)$ is the field, and $\dot{\phi}$ and $\phi'$ are its temporal and spatial derivatives. The two oscillators are described by their displacements $\xi_i(t)$, ($i=1,2$), and the corresponding heat bath variables are $q_{i\om}(t)$. We may introduce the corresponding canonical momenta like in section \ref{T1} and express the corresponding parts of the Hamiltonian directly in terms of the velocities,
\bea
H_{\rm field} &=& \int dx\,\frac12\left(\dot{\phi}(t,x)^2+\phi'(t,x)^2\right),
\label{T3.2}\\
H_{\rm osc}&=&\sum_{i=1}^2\frac{m}{2}\left(\dot{\xi}_i(t)^2+\Om^2\xi_i(t)^2\right)
,\label{T3.3}\\
H_{\rm int} &=&  e\sum_{i=1}^2\xi_i(t)\phi(t,a_i).
\label{T3.4}\eea
Further, there is a Hamiltonian for the bath fields and their interaction with the oscillators. In parallel to section \ref{T1}, it can be shown that these give only the contribution of the bath to the complete energy, which we are not interested in. This way, the thermal averages of \Ref{T3.2}-\Ref{T3.4} give the complete energy,
\be E=E_{\rm field}+E_{\rm osc}+E_{\rm int}
\label{T3.4a}\ee
with %
\be E_{\rm field}=\langle H_{\rm field} \rangle,\quad
E_{\rm osc}=\langle H_{\rm osc} \rangle,\quad E_{\rm int}=\langle H_{\rm int} \rangle.
\label{T3.5}\ee
Next we consider the equation of motion following from the Lagrangian \Ref{T3.1}. Again, like in section \ref{T1}, after eliminating the bath fields, these involve damping and Langevin forces and read
\bea    \left(\pa_t^2-\pa_x^2\right)\phi(t,x) &=&
    -e\sum_{i=1}^2 \xi_i(t)\delta(x-a_i), \nn \\
    m\left(\pa_t^2+\ga\pa_t+\Om^2\right)\xi_i(t) &=&
    -e \phi(t,a_i)+F_{i{\rm L}}(t), ~~ (i=1,2).
\label{T3.6}\eea
Here,
\be F_{i\rm L}(t)=\int_0^\infty d\om\,\frac{\mu\om^2l_0}{\sqrt{2}}
    \left(e^{-i\om t}\hat{b}_{i\om}(0)+e^{i\om t}\hat{b}^\dagger_{i\om}(0)\right)
\label{T3.7}\ee
is the Langevin force for the $i$-th oscillator and $\hat{b}_{i\om}(0)$ and $\hat{b}^\dagger_{i\om}(0)$ are the annihilation and creation operators of the heat bath attached to the $i$-th oscillator, obeying
\be [\hat{b}_{i\om}(0),\hat{b}_{j\om'}^\dagger(0)]=\delta_{ij}\delta(\om-\om')
\label{T3.8}\ee
and the thermal averages generalizing \Ref{T1.42} are
\be
\langle
\hat{b}^\dagger_{i\om}(0) \hat{b}_{j\om'}(0)
+\hat{b}_{i\om} (0)\hat{b}^\dagger_{j\om'}(0)
\rangle =
\delta_{ij}\delta(\om-\om'){\cal N}_T(\om)
\label{T3.9}\ee
with ${\cal N}_T(\om)$ given by eq. \Ref{T1.43}.

We proceed by writing down equations \Ref{T3.6} after Fourier transform \Ref{T1.27},
\bea    \left(-\om^2-\pa_x^2\right)\tphi_\om(x) &=&
    -e\sum_{i=1}^2 \txi_{i\om}\delta(x-a_i), \nn \\
    m\left(-\om^2+i\ga\om+\Om^2\right)\txi_{i\om} &=&
    -e \tphi_\om(a_i)+\tilde{F}_{i\om},\quad(i=1,2), \label{T3.10}\eea
which are generalizations of \Ref{T1.28} and of \Ref{T2.3} and \Ref{T2.4}, as well.

Like in section \ref{T2}, we proceed in two ways. Thereby, from the very beginning we consider the field $\phi(t,x)$ on a finite interval, $x\in[-\frac{L}{2},\frac{L}{2}]$, obeying Dirichlet boundary conditions \Ref{T2.29}.
\subsection{\label{secT3.1}{1$^{\rm st}$ way}}
Here we solve the second equation  \Ref{T3.10},
\be \txi_{i\om}=-\frac{e}{mN(\om)}\tphi_\om(a_i)+\frac{\tilde{F}_{i\om}}{mN(\om)},
\label{T3.12}\ee
where we took the inhomogeneous solution in line with the procedure discussed in section \ref{T1}, stating that in the presence of the heat bath the homogeneous solution dies out in time and the excitations are supported by the Langevin forces. Further, we insert solution \Ref{T3.12} into the first equation \Ref{T3.10} and come to the equation
\be   \left(-\om^2-\pa_x^2
-\frac{e^2}{mN(\om)}\sum_{i=1}^2\delta(x-a_i)\right)
\tphi_\om(x)
= -\frac{e}{mN(\om)}\sum_{i=1}^2\delta(x-a_i)\tilde{F}_{i\om}.
\label{T3.13}\ee
This equation is the inhomogeneous generalization of equation \Ref{T2.9}.
In order to solve this equation, we introduce the Green's function $G(x,x')$, obeying the equation
\be  \left(-\om^2-\pa_x^2
-\frac{e^2}{mN(\om)}\sum_{i=1}^2\delta(x-a_i)\right)
G(x,x')
 = \delta(x-x')
\label{T3.14}\ee
and the boundary conditions \Ref{T2.29}. This Green's function can be represented in the form %
\be G(x,x')
\label{T3.15} =G_0(x,x')+\sum_{r,s=1}^2
\frac{e^2}{m}G_0(x,a_r)
\Phi^{-1}_{rs}(\om)G_0(a_s,x'),
 \ee
where the 'free' Green's function $G_0(x,x')$ is given by \Ref{T2.33} and where we defined
\be \Phi_{rs}(\om)=N(\om)\delta_{rs}-\frac{e^2}{m}G_0(a_r,a_s),
\label{T3.16}\ee
which is a generalization of \Ref{T2.36}, where now
\be N(\om)=-\om^2+i\ga\om+\Om^2,
\label{T3.17}\ee
entering \Ref{T3.16},   is the same as \Ref{T1.29}, and, as before, we have $\sum_s\Phi_{rs}\Phi^{-1}_{st}=\delta_{rt}$.

We mention the following special case,
\bea G(x,a_i) &=&G_0(x,a_r)\left(\delta_{ri}+\frac{e^2}{m}\Phi^{-1}_{rs}(\om)G_0(a_s,a_i)\right)
,\nn\\ &=&
N(\om)G_0(x,a_r)\Phi^{-1}_{ri}(\om),
\label{T3.18}\eea
where \Ref{T3.16} was used. For simplifying the notation we use in these formulas and below the summation convention.

Now, using the above formulas, the solution of eq. \Ref{T3.13}, can be written down,
\be \tphi_\om(x)=-\frac{e}{mN(\om)}G(x,a_i)\tilde{F}_{i\om},
\label{T3.19a}\ee
and, using \Ref{T3.18}, in the simpler form (since expressed in terms of the 'free' Green's function),
\be \tphi_\om(x)=-\frac{e}{m}G_0(x,a_r)\Phi^{-1}_{ri}(\om)
\tilde{F}_{i\om}.
\label{T3.19}\ee
Further we insert \Ref{T3.19} into \Ref{T3.12},
\be \txi_{i\om}=
\left(\frac{e^2}{m^2N(\om)}G_0(a_i,a_r)\Phi^{-1}_{rj}(\om)+\frac{\delta_{ij}}{mN(\om)}\right)
\tilde{F}_{j\om},
\label{T3.20}\ee
which can be simplified using \Ref{T3.16},
\be \txi_{i\om}=
 \frac{1}{m}\Phi^{-1}_{ij}(\om)
\tilde{F}_{j\om}.
\label{T3.21}\ee
This way, \Ref{T3.19} and \Ref{T3.21}, are the solutions of the equations of motions obtained in the 1$^{\rm st}$ way. Both are proportional to the Langevin forces.

\subsection{\label{secT3.2}{2$^{\rm nd}$ way}}
Here we first solve eq. \Ref{T3.10},
\be \tphi_\om(x)=-e\sum_{i=1}^2G_0(x,a_i)\txi_{i\om},
\label{T3.22}\ee
where $G_0(x,x')$ is given by \Ref{T2.33}. This solution we insert into eq. \Ref{T3.10} and get %
\be %\sum_{j=1}^2
 \left(\left(-\om^2+i\ga\om+
\Om^2\right)\delta_{ij}
-\frac{e^2}{m}G_0(a_i,a_j)\right)\txi_{j\om}=\frac{1}{m}\tilde{F}_{i\om},
\label{T3.23}\ee
where, again, summation is assumed, and which by means of \Ref{T3.16}, \Ref{T3.17} can be written as
\be \Phi_{ij}(\om)\txi_{j\om}=\frac{1}{m}\tilde{F}_{i\om}
.\label{T3.24}\ee
This is an algebraic system and its solution is
\be \txi_{i\om}= \frac{1}{m}\Phi^{-1}_{ij}(\om)\tilde{F}_{j\om}.
\label{T3.25}\ee
Finally we have to insert this solution into eq. \Ref{T3.10}, which results in
\be \tphi_\om(x)=-\frac{e}{m}G_0(x,a_i)
\Phi^{-1}_{ij}(\om)\tilde{F}_{j\om}.
\label{T3.26}\ee
This way, eqs. \Ref{T3.25} and \Ref{T3.26} are the solutions of the equations of motions obtained in the 2$^{\rm nd}$ way. Both are proportional to the Langevin forces.

Comparing eqs. \Ref{T3.25} and \Ref{T3.26} with eqs. \Ref{T3.19} and \Ref{T3.21} we see, that both ways give the same expressions for the solutions. It is to be noticed, that the same expressions appear for the inhomogeneous solutions, which we consider in this subsection. The equations
\Ref{T2.9} and \Ref{T2.25} for the homogeneous solutions look different.

\subsection{\label{secT3.3}Thermal averages and free energy}
For the calculation of the free energy, we follow the procedure described in section \ref{T1}. First, we calculate the thermal averages of the Hamiltonians \Ref{T3.2}-\Ref{T3.4}, subsequently the free energy. We insert the solutions \Ref{T3.25} and \Ref{T3.26}, or equivalently,  \Ref{T3.19} and \Ref{T3.21}, and take the averages \Ref{T3.9}. Inserting  the solutions gives double integrals in $\om$, and taking the averages removes one of the integrations. Also,  double sums appear over the oscillators, which are reduced to single sums in the same way. Before starting the procedure, we mention that we by means of \Ref{T1.27}  return to the time dependent solutions.

The   procedure of taking the thermal averages can be unified for the different fields. Let
\be \hat{A}(t) = \int_{-\infty}^\infty \frac{d\om }{2\pi} e^{i\om t} a(\om) \tilde{F}_{\om},\quad
\hat{B}(t) = \int_{-\infty}^\infty \frac{d\om }{2\pi} e^{i\om t} b(\om) \tilde{F}_{\om},
\label{T3.27}\ee
be two operators like the solutions \Ref{T3.25} and \Ref{T3.26}.
Using \Ref{T1.30}, these become
\bea \hat{A}(t) &=& \int_0^\infty  {d\om }
\sqrt{\frac{\ga m \hbar \om}{\pi}}
 \left( a(-\om) e^{-i\om t}  \hat{b}_{-\om}(0)
    +a(\om) e^{-i\om t}  \hat{b}_{-\om}(0)\right),
\nn\\
\hat{B}(t) &=& \int_0^\infty  {d\om }
\sqrt{\frac{\ga m \hbar \om}{\pi}}
\label{T3.28} \left( b(-\om) e^{-i\om t}  \hat{b}_{-\om}(0)
    +b(\om) e^{-i\om t}  \hat{b}_{-\om}(0)\right).
 \eea
Now we take the thermal average of the symmetrized product of these operators and using \Ref{T3.7} and \Ref{T3.9} we get
\be
\langle \hat{A}(t)\hat{B}(t)+\hat{B}(t)\hat{A}(t)\rangle
\label{T3.29} =
\int_0^\infty d\om \,\frac{\ga m \hbar \om}{\pi}{\cal N}_T(\om)
\left(a(\om)b(-\om)+a(-\om)b(\om)\right)
.
 \ee
It should be mentioned, that all operators appear in the corresponding parts of the Hamiltonian anyway in a symmetrized way. As a consequence, the thermal averages are all real. In case the operators are equal, this statement is trivial.

In detail, we insert \Ref{T3.19} into \Ref{T3.2}, using \Ref{T3.5} and come to
\be E_{\rm field} =
\int_0^\infty d\om\,\frac{\ga m \hbar \om}{\pi}{\cal N}_T(\om)
\frac{e^2}{2m^2}\Phi^{-1}_{ij}(\om)m_{jk}
\Phi^{-1}_{ki}(-\om)
\label{T3.30}\ee
with
\be m_{jk} = \int_{-L/2}^{L/2}dx\,
\left(  \om^2 G_0(x,a_j)G_0(x,a_k)
+(\pa_x G_0(x,a_j))(\pa_x G_0(x,a_k))\right).
\label{T3.31}\ee
Further, inserting \Ref{T3.21}  into \Ref{T3.3}, using \Ref{T3.5}, one comes to
\be   E_{\rm osc}   \label{T3.32} =
\int_0^\infty d\om\,\frac{\ga m \hbar \om}{\pi}{\cal N}_T(\om)
\frac{1}{2m}(\om^2+\Om^2)
    \Phi^{-1}_{ij}(\om)\Phi^{-1}_{ji}(-\om),
 \ee
and finally, inserting  \Ref{T3.19} and  \Ref{T3.21} into \Ref{T3.3} gives
\be     E_{\rm int}  \label{T3.33} =
\int_0^\infty d\om\,\frac{\ga m \hbar \om}{\pi}{\cal N}_T(\om)
\frac{-e^2}{m^2}
    \Phi^{-1}_{ij}(\om)
    G_0(a_j,a_k)\Phi^{-1}_{ki}(-\om).
 \ee
Note, again, in all these formulas the summation convention is assumed.

Now, we investigate in more detail the structures appearing in these integrals. First we note
\be \int_{-L/2}^{L/2}dz\,G_0(x,z)G_0(y,z)=
\frac{1}{2\om}\pa_\om G_0(x,y),
\label{T3.34}\ee
which follows from \Ref{T2.33} with \Ref{T2.32}. Next we note
\be  \int_{-L/2}^{L/2}dz\,
(\pa_z G_0(x,z))(\pa_z G_0(z,y))
 =
(1+\frac{\om}{2}\pa_\om) G_0(x,y),
\label{T3.35}\ee
such that
\be m_{jk}=
(1+\om\pa_\om) G_0(a_j,a_k)
\label{T3.36}\ee
holds. We collect the relevant factors in \Ref{T3.30}, \Ref{T3.32}, \Ref{T3.33} into
\bea M &\equiv &
\frac{e^2}{2m^2}\Phi^{-1}_{ij}(\om)
    m_{jk}\Phi^{-1}_{ki}(-\om)
%\nn\\&&
+
\frac{1}{2m}(\om^2+\Om^2)
    \Phi^{-1}_{ij}(\om)\Phi^{-1}_{ji}(-\om)
\nn\\&&+
\frac{-e^2}{m^2}
    \Phi^{-1}_{ij}(\om)
    G_0(a_j,a_k)\Phi^{-1}_{ki}(-\om)
\label{T3.37}\\
&=&
\frac{1}{2m}\Phi^{-1}_{ki}(\om)
\left(  (\om^2+\Om^2)\delta_{ij}
+\frac{e^2}{m}
\left(-1+\om\pa_\om\right)G_0(a_i,a_j)
\right)
\Phi^{-1}_{jk}(-\om).
\nn\eea
Now we remember \Ref{T3.16},
\be \Phi_{rs}(\om)=(-\om^2+i\ga\om+\Om^2)\delta_{rs}-\frac{e^2}{m}G_0(a_r,a_s)
,\label{T3.38}\ee
and note its derivative,
\be \om\pa_\om\Phi_{rs}(\om)
=(-2\om^2+i\ga\om)\delta_{rs}-\frac{e^2}{m}\om\pa_\om G_0(a_r,a_s).
\label{T3.39}\ee
From these, we compose and simplify
\bea &&  \pa_\om\Phi_{il}(\om)) \Phi_{lk}(-\om)-  \Phi_{il}(\om)\pa_\om \Phi_{lk}(-\om)
\label{T3.40}\\&&~~~~~
=2i\ga((\om^2+\Om^2)\delta_{ik}+\frac{e^2}{m}(-1+\om\pa_\om)G_0(a_i,a_k)).
\nn\eea
These formulas allow  us to represent \Ref{T3.37} in the form
\bea M &=&
\frac{1}{2m}\frac{1}{2i\ga}\Phi^{-1}_{ki}(\om)
\left( 
(\pa_\om\Phi_{il}(\om)) \Phi_{lj}(-\om)-  \Phi_{il}(\om)\pa_\om \Phi_{lj}(-\om)
\right)
\Phi^{-1}_{jk}(-\om),
\nn\\&=&
\frac{1}{2m}\frac{1}{2i\ga}
\left( \pa_\om {\rm Tr}\ln \hat{\Phi}(\om)
- \pa_\om {\rm Tr}\ln \hat{\Phi}(-\om) \right),
\label{T3.41}\eea
where $\hat{\Phi}(\om)$ is the $(2\times 2)$-matrix with the entries $\Phi_{ij}(\om)$.
Also,  we used the symmetry resulting in eqs. \Ref{T2.28} and \Ref{T2.33a}.
The last line in eq. \Ref{T3.41} for   $M$ makes it meaningful to define
\be \delta(\om)=\frac{1}{2i}{\rm Tr}\ln\frac{\hat{\Phi}(\om)}{\hat{\Phi}(-\om)},
\label{T3.42}\ee
which can be viewed in parallel to \Ref{T1.60a} and \Ref{T2.17} as a kind of phase. It should be mentioned, that all three are different, their common place is the structure of being the phase of a complex expression. Confusion from using the same notation should not appear, since each is applied only within its own section of this paper.

Using \Ref{T3.42}, we write $M$ finally in the form
\be M=\frac{1}{2\ga m}\pa_\om\delta(\om).
\label{T3.43}\ee
Adding the energy, with \Ref{T3.4a} we get from \Ref{T3.30}, \Ref{T3.32}, \Ref{T3.33} and \Ref{T3.37},
\be E=
\int_0^\infty d\om\,\frac{\ga m \hbar \om}{\pi}{\cal N}_T(\om)M,
\label{T3.44}\ee
and, with \Ref{T3.43},
\be E=
\int_0^\infty \frac{d\om}{2\pi}\,
 {\hbar \om} {\cal N}_T(\om)\pa_\om\delta(\om),
\label{T3.45}\ee
which is the final expression for the thermal energy in the system of two oscillators interacting with their heat baths and with each other through the field.
Actually, it looks like eq. \Ref{T1.61} for a single oscillator interacting with a heat bath, however the meaning of the phase $\delta(\om)$ is different. Also, this formula looks like \Ref{T2.18} for the vacuum energy of the   field interacting with the oscillators, where $\delta(\om)$ has the meaning of a scattering phase shift.

The corresponding free energy can  be obtained from \Ref{T3.45} using \Ref{T1.39} and it reads
\be F=\int_0^\infty \frac{d\om}{\pi}\,
 \left(\frac{\hbar\om}{2}+T\ln\left(1-e^{-\beta\hbar\om}\right)\right)
 \pa_\om\delta(\om)
,
\label{T3.46}\ee
where the separation into zero temperature and temperature dependent parts is obvious. This formula is in parallel to \Ref{T1.62} in section \ref{T1}. Assuming $\delta(0)=0$ holds, we can integrate by parts and come to the formulas
\be F_0 = -\frac{\hbar}{2}\int_0^\infty\frac{d\om}{\pi}\,
\delta(\om),
\quad
\Delta_T F =-\hbar \int_0^\infty \frac{d\om}{\pi}\,\frac{1}{e^{\beta\hbar\om}-1}\,\delta(\om),
\label{T3.47}\ee
which are also in parallel to section \ref{T1}, eq. \Ref{T1.63b}. Also, we repeat the comment that by eqs. \Ref{T3.45}, \Ref{T3.46}, \Ref{T3.47}, the internal and free energies of the complete system including dissipation are expressed in terms of real frequencies.

We conclude this subsection by calculating the trace in $\delta(\om)$, \Ref{T3.42}. This is easily  possible since we consider the two oscillators at symmetric locations \Ref{T2.28} within the 'finite box', introduced in section \ref{secT2.2} and of equal coupling $e$. In that case the matrix $\hat{\Phi}(\om)$, whose entries $\Phi_{rs}(\om)$ are given by \Ref{T3.38}, has a particular form,
\be \hat{\Phi}=\left(-\om^2+i\ga\om+\Om^2-\frac{e^2}{m}G_1(\om)\right)
\mathbbm{1}-\frac{e^2}{m}G_2(\om)\sigma_1,
\label{T3.48}\ee
where $\mathbbm{1}$ is the unit matrix and $\sigma_1=\left(\begin{array}{cc}0&1 \\ 1&0\end{array}\right)$.
The functions $G_{1,2}(\om)$ are defined in \Ref{T2.33a}. With these formulas we have
\be {\rm Tr}\ln\hat{\Phi}(\om)
=\sum_{\sigma=\pm1}\ln\left(
\Phi_\sigma(\om)\right)
\label{T3.49}\ee
with
\be \Phi_\sigma(\om)\equiv -\om^2+i\ga\om+\Om^2-\frac{e^2}{m}G_\sigma(\om)
\label{T3.50}\ee
and
\be G_\sigma(\om)\equiv G_1(\om)+\sigma G_2(\om).
\label{T3.51}\ee
eq. \Ref{T3.49} follows since the matrixes here are $(2\times 2)$. The interpretation is that $\sigma=1$ corresponds to a symmetric function $\phi(t,x)$ and $\sigma=-1$ corresponds to an antisymmetric one. It is obvious that due to the symmetry the energy becomes a sum of the two cases.

This way, from the above formulas, for $\delta(\om)$, \Ref{T3.42}, the representation
\be \delta(\om)=\frac{1}{2i}\sum_{\sigma=\pm1}
\ln\frac{\Phi_\sigma(\om)}{\Phi_\sigma(-\om)}
\label{T3.52}\ee
holds, which we will use in the following.

Equation \Ref{T3.46} is the main result of this section. For the considered system \Ref{T3.1} it represents the free energy in case of dissipation in terms of real frequencies. In the next section we will investigate some properties of this representation.

\section{\label{T4}Special cases and Matsubara representation for the complete system in a box}
In this section, we consider in more detail the properties of the representation \Ref{T3.46} for the free energy of the complete system \Ref{T3.1}. We remind, that the system is still placed in a finite box of size $L$.

\subsection{\label{secT4.1}The limit $\ga\to0$}
First, we consider the case $\ga\to0$, i.e., the limit of vanishing dissipation.
In the energy $E$, \Ref{T3.45}, with $\delta(\om)$ given by \Ref{T3.52}, we cannot put $\ga=0$ directly in it, because the zeros of $\phi_\sigma(\om)$ and from $\phi_\sigma(-\om)$ close up on the real $\om$-axis and would quench the integration path. Instead, we are going to apply the Sokhotski-Plemelj theorem. For that we define
\be \Phi_\sigma^{(0)}(\om)=\Phi_\sigma(\om)_{|_{\ga=0}}
=-\om^2+\Om^2-\frac{e^2}{m}G_\sigma(\om).
\label{T3.53}\ee
Further we note
\be \pa_\om\ln\Phi_\sigma(\om)=\frac{\pa_\om \Phi_\sigma(\om)}{\Phi_\sigma(\om)}.
\label{T3.54}\ee
With the mentioned theorem, we have
\be\frac{1}{\Phi_\sigma(\om)}
\raisebox{-4pt}{$=\atop\ga\to 0$}
-i\pi\delta(\Phi^{(0)}_\sigma(\om) ),
\label{T3.55}\ee
(the value principal parts all cancel)
and
\be \pa_\om\ln\Phi_\sigma(\om)=
 -i\pi\delta(\Phi^{(0)}_\sigma(\om) ){\pa_\om \Phi^{(0)}_\sigma(\om)}
\label{T3.54a}\ee
holds, which results in
\be \pa_\om\delta(\om)=
 -\pi\delta(\Phi^{(0)}_\sigma(\om) ){\pa_\om \Phi^{(0)}_\sigma(\om)} .
\label{T3.55a}\ee
Inserting this expression into \Ref{T3.45}, we get for the energy
\be E
\raisebox{-4pt}{$=\atop \ga\to 0$}
-\frac12\int_0^\infty d\om\,\hbar\om
{\cal N}_T(\om)
\delta(\Phi^{(0)}_\sigma(\om) ){\pa_\om \Phi^{(0)}_\sigma(\om)}.
\label{T3.56}\ee
The function $\Phi_\sigma^{(0)}(\om)$, \Ref{T3.3}, describes the system of two oscillators interacting through the field without any heat bath or temperature. Its zeros $\om_s$,
\be \Phi_\sigma^{(0)}(\om_s)=0,
\label{T3.57}\ee
are the eigenfrequencies as defined in \Ref{T2.37}, where we, however, did not split into symmetric and antisymmetric solutions. The zeros \Ref{T3.57} can be used the rewrite the delta function in \Ref{T3.56}, and we get
\be E=\frac12\sum_s\,\hbar\om_s
{\cal N}_T(\om)
{\rm sign}(-{\pa_\om \Phi^{(0)}_\sigma(\om)_{|{\om=\om_s}}}),
\label{T3.58}\ee
which is up to the sign factor the same as \Ref{T2.38} (we dropped the regularization here). The sign factor appears somehow unexpectedly. However, for the time being, we can use a simple workaround. From \Ref{T3.53}, we note
\be \pa_\om {\Phi^{(0)}_\sigma(\om)}=
-2\om-\frac{e^2}{m}\pa_\om G_\sigma(\om),
\label{T3.59}\ee
which on $\om=\om_s$ is negative at least for small coupling, restoring the expected sign in \Ref{T3.58}. This way, we reproduce for $\ga\to 0$ the representation
\Ref{T2.38} of $E_0$ in section \ref{T2}, generalized to finite temperature by $Z=\sum_s \exp(-\beta \om_s)$, like in \Ref{T1.36}.  A similar statement holds for the free energy.

\subsection{\label{secT4.2}The limit $T\to0$}
Here we consider $T\to  0$ in the temperature dependent part $\Delta_T F$, \Ref{T3.47},  of the free energy. For this we need $\delta(\om)$, \Ref{T3.52}, for $\om\to 0$. Using \Ref{T3.50}, we have
\be \delta(\om)
\raisebox{-4pt}{$=\atop \om\to 0$}
\sum_{\sigma=\pm1}
\frac{\ga \om}{\Om^2-e^2G_\sigma(0)},
\label{T3.60}\ee
which comes in place of \Ref{T1.67} and acting the same way as in subsection \ref{secT1.5} we get
\be \Delta_T F
\raisebox{-4pt}{$=\atop T\to 0$}
-\sum_{\sigma=\pm1}  \frac{\ga T^2}{\hbar(\Om^2-\frac{e^2}{m}G_\sigma(0))}
\frac{\zeta(2)}{\pi},
\label{T3.61}\ee
from which the corresponding result for the entropy follows as usual.

We mention that we have from \Ref{T2.33a}
\be G_-(0)=\frac{b(L-b)}{2L},\quad G_+(0)=\frac{L-b}{2},
\label{T3.61a}\ee
and in \Ref{T3.60} we assumed $b<L<L_*$, with $L_*=b+(2m/b)\Om^2$, thus being in a range that the denominator does not change sign. Larger $L$ will be considered below.

\subsection{\label{secT4.3}Transition to Matsubara representation}
The transition to Matsubara frequencies  in the free energy \Ref{T3.46} can be done by turning the frequency integration towards the imaginary axis. For this, we split the logarithm in \Ref{T3.52} into two parts,
\be \delta(\om)=
\frac{1}{2i}\sum_{\sigma=\pm 1}\left(\ln \Phi_\sigma(\om)- \ln \Phi_\sigma(-\om)\right).
\label{T3.61b}\ee
Since the zeros of $\Phi_\sigma(\om)$, \Ref{T3.50}, are in the upper half plane, we turn the integration path downwards, $\om\to -i\xi$. Accordingly, in the second term we turn upwards, $\om\to i \xi$.
This choice of directions is determined by the sign of $\ga$. The Green's functions \Ref{T2.33a}, entering $\Phi_\sigma(\om)$, are defined in the box and can be continued to both sides of the real $\om$-axis. It is only after the limit $L\to\infty$, that we have to take a definite sign for $\Im \om$;  in section \ref{T2}, we took $\Im \om>0$. To some extent, one may call this procedure a kind of 'double Wick rotation', since we cannot rotate just in one direction, but we are forced to split the free energy into two parts and rotate differently in these.
%In the next section we  will do that and we will be forced to assume $\Im \om<0$ ju

Further, we use in \Ref{T3.46}
\be \frac{\hbar\om}{2}+T\ln\left(1-e^{-\beta\om}\right)=T\ln\left(2\sinh\frac{\beta\hbar\om}{2}\right)
\label{T3.61c}\ee
and its continuation
\[ \ln\left(2\sinh\frac{\beta\hbar(-i\xi)}{2}\right)=\ln\left| 2\sin\frac{\beta\hbar\xi}{2}\right |  -i\pi
\sump \Theta(\xi-\xi_l),
\]
where $\xi_l=2\pi T l$ are the Matsubara frequencies. This relation follows since the logarithm has cuts starting in $\xi=\xi_l$. This way, we get
\be F=  T \int_0^\infty d\xi\,\sump \Theta(\xi-\xi_l)
 \pa_\xi     \ln \Phi_\sigma(-i\xi_l),
\label{T3.62}\ee
where we used $\Phi_\sigma(\om)_{|_{\om=-i\xi}}=\Phi_\sigma(-\om)_{|_{\om=i\xi}}$, i.e., after the rotations the two expressions in the parenthesises in \Ref{T3.61b} become equal.

Finally, we integrate by parts and  arrive at
\be F=  T {\sum\limits_{l=0}^{\infty}}{\vphantom{\sum}}^{\prime}
\sum_{\sigma=\pm1}\ln \Phi_\sigma(-i\xi_l),
\label{T3.62a}\ee
for the free energy in terms of Matsubara frequencies.  Explicit formulas for the entries are
\be \Phi_\sigma(-i\xi_l)=\xi_l^2+\xi_l \ga+\Om^2-\frac{e^2}{m}G_\sigma(-i\xi_l)
\label{T3.63}\ee
and, from \Ref{T2.33a} and \Ref{T3.53},
\be  G_\sigma(-i\xi)\label{T3.64}
=\frac{\cosh(\xi L)-\cosh(\xi b)+\sigma(1-\cosh(\xi(L-b)))}{2\xi\sinh(\xi L)}.
 \ee
We mention, that all these expressions are real as long as the size $L$ of the box is not too large, $L<L_*$. Also, there is no problem with the zeroth Matsubara frequency, as can be seen from \Ref{T3.50} and \Ref{T3.61a}.

\section{\label{T5}Special cases and Matsubara representation for the complete system for $L\to\infty$}
In this section we consider the system of two oscillators, interacting with the field and with their heat baths, in the whole axis, i.e., we take the limit $L\to\infty$ in the final formulas of section \ref{T3}.
Actually, the box enters the formulas for the free energy \Ref{T3.46} through the 'free' Green's functions $G_{1,2}(\om)$, \Ref{T2.33a}.

Before taking the limit $L\to\infty$ of the boxes, we mention that these Green's functions enter the free energy through $\Phi_\sigma(\om)$, \Ref{T3.50}, and that we are going to make the analytic continuation into the lower half plane. As mentioned in subsection \ref{secT4.3}, this direction was determined by the circumstance that the zeros of $\Phi_\sigma(\om)$ are in the upper half plane.  Therefore,  we need to do the limit in \Ref{T2.33a} for $\Im(\om)<0$, which gives
\bea G_1^{(\infty)}(\om)&\equiv& \lim_{L\to\infty}G_1(\om)=\frac{1}{2i\om} ,  \nn\\
    G_2^{(\infty)}(\om)&\equiv& \lim_{L\to\infty}G_2(\om)=\frac{e^{-i\om b}}{2i\om}.
\label{T5.1}\eea
Inserting these into \Ref{T3.50} gives the function,
\be \Phi_\sigma^{(\infty)}(\om)=-\om^2+i\ga\om+\Om^2-\frac{e^2}{m}\frac{1+\sigma e^{-i\om b}}{2i\om},
\label{T5.2}\ee
which must be inserted into the phase \Ref{T3.52} in place of $\Phi_\sigma (\om)$, and, further, into the free energy \Ref{T3.46}.

In the following we are interested in the separation dependent part of the free energy only. It can be obtained by writing $\Phi_\sigma^{(\infty)}(\om)$ in the form
\be \Phi_\sigma^{(\infty)}(\om)=\left(-\om^2+i\ga\om+\Om^2-\frac{e^2}{2i\om m}\right)L_\sigma(\om)
\label{T5.3}\ee
with
\be L_\sigma(\om)=1-\sigma\frac{(e^2/m)\ e^{-i\om b}}{2i\om(-\om^2+i\ga\om+\Om^2)-e^2/m}.
\label{T5.4}\ee
The first factor gives (in the logarithm) the separation independent contribution and we drop it. The second factor defines a new phase,
\be \delta_L(\om)=\frac{1}{2i}\ln\frac{L(\om)}{L(-\om)},
\label{T5.5}\ee
with
\bea L(\om) &\equiv&L_+(\om)L_-(\om)
\label{T5.6} \\&=&1-\left(\frac{(e^2/m)\  e^{-i\om b}}{2i\om(-\om^2+i\ga\om+\Om^2)-e^2/m}\right)^2,
\nn\eea
where we performed the sum over $\sigma$.
This way, the separation independent part of the free energy is now
\be F=\int_0^\infty \frac{d\om}{\pi}\,
 \left(\frac{\hbar\om}{2}+T\ln\left(1-e^{-\beta\hbar\om}\right)\right)
 \pa_\om\delta_L(\om),
\label{T5.7}\ee
which is a quite explicit expression.

\subsection{\label{secT5.1}The limit $T\to0$}
We proceed in the same way as in subsection \ref{secT1.5} or \ref{secT4.2} and consider the derivative of the phase for $\om\to0$,
\be \pa_\om\delta_L(\om)= c_2+O(\om^2) ,
\label{T5.8}\ee
with
\be c_2=\frac{b^2(e^2/m)^2+2(\ga-2b\Om^2)(e^2/m)+6\Om^4}{(2\Om^2-b(e^2/m))(e^2/m)}.
\label{T5.9}\ee
The free energy becomes for $T\to0$
\be \Delta_T F       \raisebox{-4pt}{$=\atop T\to0$}
    -\frac{T^2\zeta(2)}{\pi\hbar}c_2+\dots \,.
\label{T5.10}\ee
We mention also, that for $\Om=0$ we have
\be {c_2}_{|_{\Om=0}} = -b-\frac{2 \ga}{b(e^2/m)}
\label{T5.11}\ee
and the low-T behavior of $\Delta_T F   $ does not change.

\subsection{\label{secT5.2}The transition to Matsubara representation}
The transition to Matsubara frequencies can be done in much the same way as in subsection \ref{secT4.3}. We split the phase,
\be \delta_L(\om)=
\frac{1}{2i} \left(\ln L(\om)- \ln L(-\om)\right).
\label{T5.12}\ee
into two parts and turn the contour downwards, $\om=-i\xi$, in the first part, and upwards, $\om=i\xi$, in the second part. As  already mentioned, this can be viewed as a kind of double Wick rotation. However, in doing so we write the integral over $\om$ in the free energy \Ref{T5.7}, which is an integral over the difference in \Ref{T5.12}, as a difference of integrals. This is not directly possible since now for $\om\to0$
\be L(\om)=2i\om\left(b-\frac{2\Om^2}{e^2/m}\right)+O(\om^2)
\label{T5.13}\ee
holds, as can be seen easily from \Ref{T5.6}. As a consequence, each of the integrals would diverge at $\om=0$. The obvious way out is to integrate over $\om$ first from some $\ep>0$ and to consider the limit $\ep\to0$ after the rotations,
\bea F  &=&\frac{1}{2\pi i}\lim_{\ep\to0}
     \left[\int_\ep^\infty d\om\, \left(\frac{\hbar\om}{2}+T\ln\left(1-e^{-\beta\hbar\om}\right)\right) \pa_\om \ln L(\om)
 \right.\nn\\&& \left.   - \int_\ep^\infty d\om\,\left(\frac{\hbar\om}{2}+T\ln\left(1-e^{-\beta\hbar\om}\right)\right) \pa_\om \ln L(-\om)\right].
\label{T5.14}\eea
Now we split the integration pathes   into two parts each. The first part is a half circle of radius $\ep$, given by $\om=\ep e^{-i\varphi}$ in the first integral and $\om=\ep e^{i\varphi}$ in the second one,
with $\varphi=0 \dots \frac{\pi}{2}$ in both cases. The second part is a straight line along the imaginary axis with  $\om=-i\xi$  in the first integral and $\om=i\xi$  in the second one, with $\xi\in [\ep,\infty)$ in both cases.
From the integrals along the straight lines, using \Ref{T3.61c},  we have
\bea &&F_{\rm lin.}=\frac{T}{2\pi i}\label{T5.15}\\&& \left[ \int_\ep^\infty d\xi\,
    \left( \ln \left|2\sin\frac{\beta\xi}{2}\right|-i\pi\sump\Theta(\xi-\xi_l)\right)
            \pa_\xi\left(\ln\left|L(-i\xi)\right|+i\pi\Theta(\xi_*-\xi)\right)
     \right.\nn\\&& \left.
   -\int_\ep^\infty d\xi\,  \left( \ln \left|2\sin\frac{\beta\xi}{2}\right|+i\pi\sump\Theta(\xi-\xi_l)\right)
            \pa_\xi\left(\ln\left|L(-i\xi)\right|-i\pi\Theta(\xi_*-\xi)\right) \right].
\nn\eea
An additional feature appeared here since the function $L(-i\xi)$ has a zero, i.e., there is a real $\xi_*>0$ such that $L(-i\xi_*)=0$ holds. For this reason the logarithms of $L$ acquire imaginary parts as shown in the above formula. The signs follow from $L(-i\xi)>0$ for sufficiently large $\xi$ and the side on which the branch point of the logarithm is passed. The origin of these zeros are the bound states mentioned earlier. Also, we mention that this feature appears  within the finite box for $L>L_*$ as well.

Simplifying and carrying out the integration in the contribution with $l=0$,   accounting for $\Theta(\xi-\xi_0)=1$ for $\xi>0$, we get
\be F_{\rm lin.}=\frac{T}{2} \ln\left|L(-i\ep)\right|+T\sum_{l=1}^\infty \ln\left|L(-i\xi_l)\right|
    -T \left|2\sin\frac{\beta\hbar\xi_*}{2}\right|.
\label{T5.16}\ee
Indeed, this expression is divergent for $\ep\to0$. Using \Ref{T5.13},
\be \frac{T}{2} \ln\left|L(-i\ep)\right|
    =\frac{T}{2}\ln\ep+\frac{T}{2}\ln\left|2\left(b-\frac{2\Om^2}{e^2/m}\right)\right|+O(\ep)
\label{T5.17}\ee
holds and we see that the divergence is logarithmic.

It remains to calculate the contribution from the half circles.  Restricting to contributions not vanishing for $\ep\to0$  we note
\be \frac{\hbar\om}{2}+T\ln\left(1-e^{-\beta\hbar\om}\right)\sim
    \ln(\beta\hbar\ep e^{-i\varphi})
\nn\ee
in the first contribution and the complex conjugate of that in the second. With \Ref{T5.13} we note $\pa_\xi\ln L(-i\xi)=\frac{1}{\xi}+O(1)$. Inserting into \Ref{T5.7} gives
\bea F_{\rm h.circ.} &=& -\frac{T}{\pi}\int_0^{\pi/2} d\varphi\,
    \left(\ln(\beta\hbar\ep e^{-i\varphi})+\ln(\beta\hbar\ep e^{i\varphi})\right),
\nn\\&=&    -\frac{T}{2}\ln(\beta\hbar\ep)+O(\ep).
\label{T5.18}\eea
We see, the divergence for $\ep\to0$ compensates just that in \Ref{T5.16}, as expected, and we get for the sum of \Ref{T5.16} and \Ref{T5.18}
\be  F = \frac{T}{2}\ln\left|2\left(b-\frac{2\Om^2}{e^2/m}\right)T\right|
    +T\sum_{l=1}^\infty \ln\left|L(-i\xi_l)\right|
    -T \left|2\sin\frac{\beta\hbar\xi_*}{2}\right|.
\label{T5.19}\ee
This formula, together with
\be L(-i\xi)=1-\left(\frac{(e^2/m)\  e^{-\xi b}}{2\xi(\xi^2+\ga\xi+\Om^2)-e^2/m}\right)^2,
\label{T5.20}\ee
represent the final expressions for the free energy in case of dissipation in Matsubara representation.
The difference to \Ref{T3.62a} can be formulated as formal substitution,
\be L(-i\xi_{l=0}) \to 2\left(b-\frac{2\Om^2}{e^2/m}\right)T,
\label{T5.21}\ee
in the contribution from the zeroth Matsubara frequency. It should be mentioned that this modification does not depend on dissipation ($\ga$ does not appear in \Ref{T5.21}), but is rather a property of the model.

\section{\label{concl}Conclusions}
We considered a simple system, consisting of two harmonic oscillators, interacting with a scalar field in (1+1)-dimensions and with heat baths, as  model for the Casimir-Polder force in electrodynamics. The interaction with the heat baths gives the system a dissipation, introduced from 'first principles' in the sense, that this system has a Hamiltonian. First, we consider the system without a heat bath and conclude, that it is inevitable first to use a finite box in order the get a meaningful description.
Next, we include the interaction with the heat bath. Here, we derive a quite simple representation of the free energy in terms of real frequencies, eq. \Ref{T3.46} with \Ref{T3.52} and \Ref{T3.50}, which we   did not find in literature with such simple and direct derivation. However, it is yet a generalization of the 'remarkable formula' found in \cite{ford85-55-2273}. We mention  that these real frequencies are not the eigenfrequencies (resonances) of the dissipative system. We investigated the special cases like vanishing dissipation and showed that the results reproduce the case taken without dissipation from the beginning.

Specifically,  we consider the low temperature behavior and show that no thermodynamic problem occurs. This holds also in the case of vanishing intrinsic frequency $\Om$ of the oscillators.
All that is done for the system in a box. Then, we consider the case $L\to\infty$, i.e., of removing the box. A quite explicit formula for the free energy, eq. \Ref{T5.7}, with \Ref{T5.5} and \Ref{T5.6} inserted,  appears.
In this case, the system always has a critical behavior, which can be inferred easily from the delta functions in eq. \Ref{T2.7}, which are attractive and result in bound states of the $\phi$-field resp. in the zeros of the function $L(-i\xi)$ in \Ref{T5.20}. It is only in a large separation expansion that these do not show up, making the expansion an asymptotic one. It is also to be mentioned that the critical behavior appears already in a finite box, if it is sufficiently large.

It is interesting to mention, that also for the size of the box exceeding some critical value, one or two bound states for the $\phi$-field appear in the considered system. We assume that the corresponding states are occupied. As a result, an additional contribution in the Matsubara representation, the last term in F, \Ref{T5.19}, appears.

Finally, we consider the Matsubara representation for the free energy on the whole axis. We start from our representation in terms of real frequencies and make a kind of double Wick rotation. We obtain the expected contributions  from the Matsubara frequencies $\xi_l$ with $l\ge 1$, and, as a new feature, a modification of the contribution from the zeroth Matsubara frequency. It can be formulated in terms of the formal rule \Ref{T5.21}. Besides the representation of the free energy in terms of a phase, this is a main result of the present paper.

We mention that the transition to vanishing dissipation ($\ga\to0$) is smooth even in the Matsu\-bara representation. This can be seen from $L(-i\xi)$ in \Ref{T5.20} which gives the same result to the free energy for $\ga\to0$ as putting $\ga=0$ in from the very beginning as  in the plasma model, eq. \Ref{T2.22} for $t(i\xi)^{-1}$. The reason for this is the presence of the $(-e^2/m)$-term in the denominator  of \Ref{T5.20}, which can be traced back to $G_1$ from \Ref{T5.1} resp. in \Ref{T3.48}, which is a result of the self field of an oscillator, which is finite in $(1+1)$-dimensions. If we had discarded this self field contribution as we are forced to do in higher dimensions \cite{bord15-91-065027}, then the limit $\ga\to0$ would not reproduce the the free energy of the  plasma model.

Obvious future work is the application of the me\-thods developed in this paper to the Casimir-Polder force in electrodynamics.

\section*{Acknowledgement}I would like to ask the referee for very careful reading of the manuscript.

\end{document}